\documentclass[fontsize=11pt,a4paper,fleqn]{scrartcl}

\usepackage{a4}
\usepackage{layouts}
\usepackage{setspace}

\usepackage{amsfonts}
\usepackage{amsmath}
\usepackage{amssymb}
\usepackage{exscale} 

\usepackage[pdftex]{graphicx} \usepackage[all]{xy}
\usepackage{tikz}
\usetikzlibrary{arrows,matrix,backgrounds,shapes,positioning,fit,chains}
\usepackage{rotating}
\usepackage{caption}
\usepackage{subfig}
\usepackage[noend]{algorithmic}
\usepackage{algorithm}

\usepackage{multirow}

\usepackage{enumerate}

\usepackage[english]{babel} \usepackage[T1]{fontenc}
\usepackage[ansinew]{inputenc}
\usepackage{html}
\usepackage{url}

\usepackage{comment}
\usepackage{color}
\usepackage[colorinlistoftodos]{todonotes}

\usepackage{makeidx}
\usepackage{natbib}

 \usepackage[sc]{mathpazo}
\setkomafont{sectioning}{\usefont{OT1}{cmss}{c}{n}\selectfont}
\setkomafont{title}{\usefont{OT1}{cmss}{c}{n}\selectfont \huge} \setkomafont{section}{\LARGE} \setkomafont{subsection}{\Large} \setkomafont{subsubsection}{\large\itshape}
 \allowdisplaybreaks[4]

\newcommand{\tr}{^\prime}
\newcommand{\bc}{\begin{center}}
\newcommand{\ec}{\end{center}}
\newcommand{\ben}{\begin{enumerate}}
\newcommand{\een}{\end{enumerate}}
\newcommand{\beq}{\begin{equation*}}
\newcommand{\eeq}{\end{equation*}}
\newcommand{\bea}{\begin{align*}}
\newcommand{\eea}{\end{align*}}
\newcommand{\bi}{\begin{itemize}}
\newcommand{\ei}{\end{itemize}}

\newcommand{\bm}{\boldsymbol}
\newcommand{\bvec}{\left(\begin{array}{c}}
\newcommand{\evec}{\end{array}\right)}
\newcommand{\bmat}[1]{\left(\begin{array}{*{#1}{c}}}
\newcommand{\emat}{\end{array}\right)}

\newcommand{\blind}[1]{#1}

\newcommand{\iid} {\operatorname{i.i.d.}}

\newcommand{\package}[1]{\textsf{#1}}
\newcommand{\data}[1]{\texttt{#1}}

\algsetup{linenodelimiter=}

\title{\vspace*{-6cm}Spike-and-Slab Priors for Function Selection in Structured Additive Regression Models}
\author{\blind{Fabian Scheipl}, \blind{Ludwig Fahrmeir}, \blind{Thomas Kneib}
\thanks{\blind{Fabian Scheipl is Postdoctoral Fellow (E-mail:
fabian.scheipl@stat.uni-muenchen.de) and Ludwig Fahrmeir is Professor (emeritus), Department of Statistics, Ludwig-Maximilians-Universit\"at M\"unchen, Munich, Germany.
Thomas Kneib is Professor, Department of Economics, Georg-August-Universität Göttingen, Göttingen, Germany. This work was supported by the German Science Foundation (DFG grant FA
128/5-1).}}}
\date{}

\begin{document}

 \maketitle
\vspace*{-4em}
\begin{abstract}
Structured additive regression provides a general framework for complex Gaussian and non-Gaussian regression models, with predictors comprising arbitrary combinations of nonlinear
functions and surfaces, spatial effects, varying coefficients, random effects and further regression terms. The large flexibility of structured additive regression makes function
selection a challenging and important task, aiming at (1) selecting the relevant covariates, (2) choosing an appropriate and parsimonious representation of the impact of covariates
on the predictor and (3) determining the required interactions. We propose a spike-and-slab prior structure for function selection that allows to include or exclude single
coefficients as well as blocks of coefficients representing specific model terms. A novel multiplicative parameter expansion is required to obtain good mixing and convergence
properties in a Markov chain Monte Carlo simulation approach and is shown to induce desirable shrinkage properties. In simulation studies and with (real) benchmark classification
data, we investigate sensitivity to hyperparameter settings and compare performance to competitors. The flexibility and applicability of our approach are demonstrated in an additive
piecewise exponential model with time-varying effects for right-censored survival times of intensive care patients with sepsis. Geoadditive and additive mixed logit model
applications are discussed in an extensive appendix.
\end{abstract}
\vspace*{-1em} \textit{Key-words: parameter expansion, penalized splines, stochastic search variable selection, generalized additive mixed models, spatial regression}

\section{INTRODUCTION}

Recent research on function selection mostly considers the additive model \linebreak \mbox{$y=f_1(x_1)+\ldots+f_q(x_q)+\epsilon$} for Gaussian responses, sometimes including
additional linear effects or interactions of functions. In this paper we introduce a spike-and-slab prior structure to perform Bayesian inference and function selection in structured
additive regression (STAR) models, i.e., in exponential family regression models with additive predictors incorporating different types of functions or effects. Compared to additive
models, we therefore extend function selection to regression with non-Gaussian, in particular discrete responses and the predictor may contain additional components, such as varying
coefficient terms $uf(x)$, smooth interactions $f(x_1,x_2)$, spatial effects $f_{\operatorname{geo}}(s)$ for geoadditive regression, and cluster-specific random effects. Functions of
continuous covariates are represented through penalized (tensor product) splines, $f_{\operatorname{geo}}(s)$ through (conditionally) Gaussian Markov random fields, and
cluster-specific effects through (conditionally) Gaussian i.i.d. random effects, but other basic function expansions, surface smoothers, and spatial models are possible as well, see
e.g., \citet{Fahr:Kneib:Lang:2004} for details. Any generalized structured additive regression model can then be written in unifying form as
\begin{equation}\label{eq:genpred:matrix}
 E(\bm y|\bm\eta) = h(\bm \eta), \qquad \bm \eta = \bm f_1 + \ldots + \bm f_P =
 \bm Z_1\bm\delta_1 + \ldots + \bm Z_P\bm\delta_P,
\end{equation}
where the conditional expectation $E(\bm y|\bm\eta)$ of the response vector $\bm y=(y_1,\ldots,y_n)'$ is related to a predictor vector $\bm \eta= (\eta_1,\ldots,\eta_n)'$ via a known
response function $h$ as in generalized linear models. The predictor vector $\bm \eta$ is additively composed of covariate effects $\bm f_j= (f_j(\bm x_{j1}), \ldots, f_j(\bm
x_{jn}))'$ of different types. Each effect $\bm f_j$ can also be a function of multiple covariates and is represented by suitable design matrices $\bm Z_j$ of dimension $(n \times
D_j)$, and $D_j$-dimensional regression coefficients
 $\bm\delta_j|s_j^2 \sim N(\bm 0, s_j^2 \bm P_j^{-})$
with fixed positive (semi-)definite scaled precision matrix $\bm P_j$ and prior variance $s_j^2$. Singular precision matrices result from Bayesian P-Splines \citep{Brezger:Lang:2006}
or intrinsic Markov random fields \citep{Rue:Held:2005}.

Section~\ref{sec:methods:reparam} shows how the terms in~\eqref{eq:genpred:matrix} can be reparameterized in terms of modified design matrices associated with conditionally Gaussian
i.i.d. coefficient vectors. 
This reparameterization essentially follows ideas similar to those considered in \citet{RupWanCar03} or \citet{Fahr:Kneib:Lang:2004} but is especially designed to improve the
selection properties of our approach. It also has the advantage that additional function selection questions can be handled, for example differentiating between no effect, linear
effect and inherently nonlinear effect of a continuous covariate (see Section~\ref{sec:methods:reparam} for details).

In function selection, we are interested in finding simple special cases of \eqref{eq:genpred:matrix}, where some of the functions are identified as having negligible impact on the
response. For example, in our geoadditive regression model of rents in the city of Munich, we want to select a subset from a large set of categorical covariates and to decide if
effects of age of a building and floor space of the flat are nonlinear or linear, if an interaction between them is necessary, and if a spatial effect in form of a Markov random
field representing the location of buildings is needed. In the application dealing with the analysis of survival times of patients that acquired a septic infection after surgery, we
are interested in finding a parsimonious model to indicate which covariates have (potentially nonlinear) impact on survival and to evaluate the presence or absence of time-varying
effects indicating non-proportional hazards.

In a Bayesian or mixed models framework, functional effects in structured additive regression are reparameterized as i.i.d. Gaussian random effects (or i.i.d. Gaussian priors
from a Bayesian perspective). This ideas has become quite popular since it allows treating complex regression models as mixed models and makes corresponding restricted maximum
likelihood (REML) estimation available for the determination of smoothing variances. \citet{Wan00} was among the first to recognize this possibility and introduced mixed model based
inference for penalized spline regression based on truncated power series expansions in Gaussian regression models. \citet{RupWanCar03} provide an in-depth overview on mixed model
based semiparametric regression and describe extensions to the estimation of interaction surfaces and spatial effects. In an (empirical) Bayes framework, the mixed model representation
corresponds to a reparameterisation of the prior and REML estimation is interpreted as marginal likelihood estimation, see \citet{Fahr:Kneib:Lang:2004} for more details in the
context of (generalized) STAR models and \citet{CraRupWan05} for the implementation of such models via WinBUGS. The mixed model perspective on semiparametric regression has also led
to the development of likelihood ratio tests that implement the selection of functions $f_j(x_j)$ based on testing $H_0: s_j^2=0$ versus $H_A: s_j^2>0$ \citep[cf.][]{CraRupCla05,
Greven:Crainiceanu:2008, Scheipl:Greven:Kuech:2008}, since
$s_j^2=0$ implies $\bm f_j = \bm Z_j\bm\delta_j = \bm 0$. However, these likelihood ratio tests are so far only applicable in Gaussian regression models and are not suitable for automatic function selection in complex regression models with a large number of potentially nonlinear effects.

To develop a Bayesian counterpart of likelihood ratio tests, it seems natural to impose a bimodal spike-and-slab prior on the variances $s_j^2$, as suggested in \citet{Ishwaran:2005}
for the case of variable selection in high-dimensional linear models, i.e. for selecting single scalar regression coefficients. Indeed, our first attempt to select functions in STAR
models was based on this simple idea. However, as shown in the web appendix, such a straightforward approach is rendered infeasible by the severe convergence and mixing problems it
causes. Informally, the problem is that a small variance for a block of coefficients implies small coefficient values and small coefficient values in turn imply a small variance.
Therefore, blockwise MCMC samplers are unlikely to exit a basin of attraction around zero. We therefore propose a multiplicative parameter expansion for the regression coefficients
inspired by the work of \citet{Gelman:2008} in the context of mixed models. We show that this parameter expansion leads to an efficient MCMC strategy and to a prior with
regularization properties similar to $\mathcal{L}_q$-penalization with $q<1$ in Section~\ref{sec:methods:shrinkage}.

The main advantages of our new prior structure can be summarized as follows: First, unlike standard stochastic search variable selection approaches, it can routinely be used with
non-Gaussian responses from the exponential family based on iteratively weighted least squares updates. Second, it supports the full generality of STAR models, i.e. it accommodates
all types of regularized effects with a (conditionally) Gaussian prior such as simple covariates or covariate blocks (both continuous and categorical), penalized splines (uni- or
multivariate), spatial effects, random effects or ridge-penalized factors and all their interactions (e.g.~(space-)varying coefficient terms or random slopes). Third, it scales to
data sets of intermediate size with thousands of observations and high-dimensional predictors including hundreds of model terms. Fourth, it is implemented in publicly available and
user-friendly open source software (R-package \blind{\package{spikeSlabGAM}} \blind{\citep{spikeSlabGAMJSS}}), therefore allowing reproducibility of our results and immediate
application to new data sets.

Due to the practical importance of the topic, there is a vast amount of literature on selecting components in predictors of regression models. Most previous work considers selection
of variables or associated (single) regression coefficients in high-dimensional (generalized) linear models. Penalization approaches have become quite popular, in particular the
Lasso or the SCAD penalty and modifications as the adaptive or group Lasso. Another branch is boosting, see \citet{Buehlmann:Hothorn:2007} for a survey. Most Bayesian approaches
for variable selection are based on spike-and-slab priors for regression coefficients, see for example the stochastic search variable selection (SSVS) approach in
\citet{George:McCulloch:1993}, among other methods, and the review in \citet{OHara:Sillanpaa:2009}.

In comparison, research on function selection is more sparse. Recently, \citet{Marra:Wood:2011} have proposed a ``double shrinkage'' approach for GAMs with an additional penalty on
the null space of the smoothness penalty which enables shrinking entire functional terms to zero. Most other penalization methods only consider additive models for continuous
(Gaussian) responses and perform function selection by penalizing certain norms of functional components or associated blocks of basis function coefficients in Lasso- or SCAD-type
fashion. \citet{Lin:Zhang:2006} proposed the component selection and smoothing operator (COSSO) in additive smoothing spline ANOVA models. Motivated by the adaptive group Lasso,
\citet{Storlie:Bondell:Reich:2010} propose the adaptive COSSO (ACOSSO) to penalize each functional component differently so that more flexibility is obtained. Its superior
performance to the COSSO (and MARS) is demonstrated for simulated and real data. A similar adaptive group Lasso approach is studied in \citet{Huang:Horowitz:Wei:2010}.
\citet{Ravikumar:2009} estimate sparse additive models by penalizing the quadratic norm of functional predictor terms, \citet{Meier:Geer:Buehlmann:2009} additionally impose a
smoothness penalty.

Many Bayesian function selection approaches are based on introducing spike-and-slab priors with a point mass at zero directly for blocks of basis function coefficients or,
equivalently, indicator variables for functions being zero or nonzero. \citet{Wood:Kohn:2002} and \citet{Yau:Kohn:Wood:2003} describe implementations using a data-based prior that
requires two MCMC runs, a pilot run to obtain a data-based prior for the ``slab'' part and a second one to estimate parameters and select model components.
\citet{Panagiotelis:Smith:2008} combine this stochastic search variable selection approach with partially improper Gaussian priors, as for basis coefficients of Bayesian P-splines,
in high-dimensional additive models. They suggest several sampling schemes that dominate the scheme in \citet{Yau:Kohn:Wood:2003}. A more general approach based on double exponential
regression models that also allows for flexible modeling of the dispersion is described by \citet{Cottet:Kohn:Nott:2008}. They use a reduced rank representation of cubic smoothing
splines with a very small number of basis functions to model the smooth terms in order to reduce the complexity of the fitted models, and, presumably, to avoid the mixing problems
already mentioned and described in the web appendix. \citet{Reich:Storlie:Bondell:2009} use the smoothing spline ANOVA framework and a spike-and-slab prior for the variance of
Gaussian process priors to perform variable and function selection via SSVS for Gaussian responses. In a wavelet-based functional Gaussian mixed model framework, \citet{MorCar06}
place spike and slab priors directly on wavelet coefficients to decide whether they are important for representing functional effects. This approach is further extended by
\citet{ZhuBroMor11} who improve robustness and adaptivity by considering scale mixtures of normals in the spike and slab specification. \citet{Zhu:Vannucci:2010} develop a probit
model for functional data classification that allows to select functional predictors. Their hierarchical model is based on a latent Gaussian model and on Gaussian process priors for
functional effects. Like \citet{Reich:Storlie:Bondell:2009}, they place a spike and slab prior on the variance, while the remaining part of the Gaussian process covariance matrix is
assumed to be known. This (strong) assumption and the possibility to integrate out functional effect parameters in full conditionals for variance parameters facilitates MCMC
inference in this case.

Function selection in generalized additive models using Bayes factors has recently been proposed in two ways: \citet{Sabanes:Held:2011} present a basis function selection approach
for Bayesian fractional polynomials for potentially nonlinear effects, while the approach of \citet{Sabanes:Held:2011b} is based on a grid of fixed effective degrees of freedom for
each penalised spline. \citet{Chipman:George:McCulloch:2010} propose Bayesian adaptive regression trees (BART) to develop a completely non-parametric prediction-oriented approach.

In principle, adaptive smoothing approaches based on knot selection strategies as suggested for example in \citet{Denison:Mallick:Smith:1998} for the Bayesian MARS or
\citet{DimGenKas01} for adaptive regression splines (BARS) could be considered as additional competitors for function selection. In particular, knot selection strategies typically
involve the possibility to deselect covariate effects by excluding all basis functions associated with a specific covariate. However, this possibility is usually only a by-product of
the model specification and is not directly intended for function selection.  Roughly, these approaches correspond to specifying separate i.~i.~d.~ spike and slab priors for the
scalar basis function coefficients, rather than imposing a multivariate spike and slab prior on the entire vector of basis function coefficients as in
\citet{Panagiotelis:Smith:2008}. Note that even fitting of a single function with adaptive smoothing based on knot selection can be extremely time-consuming, see for example the
comparison of BARS with the adaptive penalty approach in \citet{KriCraKau08}. As a consequence, selecting functions in this fashion becomes inefficient, if not computationally
infeasible. Therefore we will not consider knot-selection approaches in the rest of this paper.

\section{NMIG PRIORS FOR FUNCTION SELECTION}\label{sec:methods}

\subsection{Generic Parameterization}\label{sec:methods:reparam}

Many of the regression terms available for STAR models are associated with conditionally Gaussian priors with zero mean and general positive semidefinite precision matrices $\bm P$
(cf.~\eqref{eq:genpred:matrix}). For example, in the case of penalized splines the precision matrix represents the dependence structure imposed by the random walk prior
\citep{Brezger:Lang:2006} or in the case of Gaussian Markov random fields, the precision matrix is defined by the neighborhood structure underlying the geographical arrangement of
the data \citep{Rue:Held:2005}. We will now show that such general Gaussian priors can always be recast based on i.i.d. priors:

Assume that $\bm\delta| s^2 \sim N(\bm 0, s^2 \bm P^-)$ represents the $D$-dimensional regression coefficient corresponding to one of the terms $f(\bm z) = \bm{Z \delta}$ appearing
in a STAR model \eqref{eq:genpred:matrix}. Let $K$ denote the dimension of the null-space of $\bm P$. Since $f(\bm z) =  \bm{Z \delta}$ and $\bm{Z \delta}|s^2 \sim N(\bm 0, s^2 \bm{
Z P^- Z}\tr)$, the spectral decomposition $\bm{Z P^-} \bm Z\tr= \bm{UVU}\tr$ with orthonormal $\bm U$ and a diagonal $\bm V$  with entries $\geq 0$ yields an orthogonal basis
representation for the improper prior covariance of $f(\bm z)$. For $\bm{Z}$ with $D$ columns and full column rank and $\bm P$ with rank $d=D - K$, all eigenvalues in $\bm V$ except
the first $d$ are zero. Now write $\bm{Z P^-} \bm Z\tr = [\bm U_+  \bm U_0]\tr \left[\begin{smallmatrix} \bm V_+ & \bm 0 \\ \bm 0 & \bm 0\end{smallmatrix}\right] [\bm U_+  \bm U_0]$,
where $\bm U_+$ is a matrix of eigenvectors associated with the positive eigenvalues in $\bm V_+$, and $\bm U_0$ are the eigenvectors associated with the zero eigenvalues. With $\bm
X_{\operatorname{pen}} = \bm U_+ \bm V_+^{1/2}$ and $\bm\beta_{\operatorname{pen}} \sim N(\bm 0, v^2\bm I)$, $f_{\operatorname{pen}}(\bm z)= \bm X_{\operatorname{pen}}
\bm\beta_{\operatorname{pen}}$ has a proper Gaussian distribution that is proportional to that of the partially improper prior of $f(\bm z)$ \citep[][eq. (3.16)]{Rue:Held:2005} but
parameterizes only the penalized component of $f(\bm z)$, while $\bm X_0 = \bm U_0$ and the associated coefficients $\bm\beta_0$ parameterize functions $f_0(\bm z) = \bm X_0
\bm\beta_0$ in the null space of $\bm P$. In summary, we reparameterize and decompose $f(\bm z) = f_0(\bm z) + f_{\operatorname{pen}}(\bm z)$. Our reparameterization follows similar
ideas as in early references on mixed model based inference in semiparametric regression (see for example \citet{Wan00,RupWanCar03} for corresponding frequentist approaches and
\citet{Fahr:Kneib:Lang:2004,CraRupWan05} for Bayesian interpretations) but is designed for the special purpose of function selection. Basing the decomposition on the spectral
decomposition of $\bm{Z P^-} \bm Z\tr$ instead of the spectral decomposition of $\bm P$ yields an orthogonal basis representation and therefore facilitates the differentiation
between penalized and unpenalized parts of a function.

In practice, it is unnecessary and impractically slow to compute all $n$ eigenvectors and values for a full spectral decomposition $\bm{UVU}\tr$. Only the first $d$ are needed for
$\bm X$, and of those the first few typically represent most of the variability in $f(\bm z)$. Our implementation makes use of a fast truncated bidiagonalization algorithm
\citep{Baglama:Reichel:2006} available in \package{irlba} \citep{irlba} to compute only the largest $d$ eigenvalues of $\operatorname{Cov}\left(f(\bm z)\right)$ and their
associated eigenvectors. Only the first $d$ eigenvectors and -values whose sum represents at least $.995$ of the sum of all eigenvalues are used to construct the reduced rank
orthogonal basis $\bm X_{\operatorname{pen}}$ with $d$ columns. E.g.~for a cubic P-spline with second order difference penalty and 20 basis functions (i.e.~$D = 20$ columns in
$\bm{Z}$ and $K=2$), $\bm X$ will often have only 8 to 12 columns and $\bm X_0$ has one column for the linear trend since the constant part of $f_0(\bm z)$ is subsumed into the
global intercept to ensure identifiability.

The advantages of applying a reparameterization are two-fold: First, we can separate the penalized part of a predictor function $f(\bm z)$. Second, we can now not only assign an
i.i.d. Gaussian prior to the penalized part but also to the unpenalized part. Of course we then no longer have a one-to-one transformation of the original prior but we can perform
function selection on both the penalized and the unpenalized parts. For example, in case of penalized splines with $k$-th order random walk prior, the space of unpenalized functions
consists of all polynomials of order less than $k-1$. Separating these polynomials from the non-polynomial, penalized part of the function opens up the possibility to decide whether
a nonlinear effect for a continuous covariate should be included in the model at all, whether it can be described in terms of a linear effect or whether a nonlinear effect is needed.
The resulting models are more parsimonious and easier to interpret.

In the following, we assume that the reparameterization has been applied to all relevant model terms and we treat $f_0(\bm z)$ and$f_{\operatorname{pen}}(\bm z)$ as separate
model terms, so that $\bm \eta = \bm\eta_0 + \sum_{j=1}^P \bm Z_j \bm\delta_j$ with $\bm \delta_j|s_j^2 \sim N(\bm 0, s_j^2 \bm P_j^-)$ (cf.~\eqref{eq:genpred:matrix}) can now be
rewritten as
\begin{align}\label{eq:genpred:matrix_reparam}
  \bm \eta &= \bm\eta_0 + \sum_{j=1}^p \bm X_j \bm\beta_j,
\end{align}
with $\bm\eta_0$ containing a global intercept, offset terms, and effects that are not associated with a variable selection prior, $\bm \beta_j|v_j^2 \sim N(\bm 0, v_j^2 \bm I)$ and
$p \geq P$ the new number of separate model terms now comprising both $\bm X_{0,j}$ and $\bm X_{\operatorname{pen},j}$, $j=1,\ldots,p$.

\subsection{Parameter-Expanded NMIG Prior}\label{sec:methods:penmig}

Inspired by the work of \citet{Gelman:2008}, we propose a multiplicative parameterization of $\bm\beta_j$ and combine it with a spike-and-slab prior based on a mixture of inverse
gamma distributions for the variances $v_j^2$. More specifically, we multiplicatively expand the $d_j$-dimensional vector $\bm\beta_j$ to $\bm \beta_j = \alpha_j \bm\xi_j$, $\bm\xi_j
\in \mathbb{R}^{d_j}$, where the scalar parameter $\alpha_j$ parameterizes the importance of the $j$-th coefficient block, while $\bm\xi_j$ ``distributes'' $\alpha_j$ across the
entries in $\bm \beta_j$.

We assume that $\alpha_j\sim N(0, v_j^2)$ follows a univariate Gaussian distribution with variance $v_j^2=\gamma_j\tau_j^2$ given by the product of an indicator variable $\gamma_j$
and the prior variance $\tau_j^2$. In a further level of the hierarchy we specify
\[
 \tau_j^2\sim\Gamma^{-1}(a_\tau, b_\tau), \quad \gamma_j\sim
 w\delta_1(\gamma_j) + (1-w)\delta_{v_0}(\gamma_j),
\]
i.e. the variance $\tau_j^2$ is assumed to follow an inverse gamma-prior with shape parameter $a_\tau$ and scale parameter $b_\tau$ chosen such that $b_\tau \gg a_\tau$. As a
consequence, the mode $b_\tau / a_\tau$ is significantly greater than 1. The indicator $\gamma_j$ takes the value 1 with probability $w$ or some (very) small value $v_0$ with
probability $1-w$. The implied prior for the effective variance $v_j^2 = \gamma_j\tau_j^2$ is a bimodal mixture of inverse gamma distributions, with one component strongly
concentrated on very small values -- the \emph{spike} with $\gamma_j=v_0$ and effective scale parameter $v_0 b_\tau$ -- and a second more diffuse component with most mass on larger
values -- the \emph{slab} with $\gamma_j=1$ and scale $b_\tau$. A coefficient associated with a variance that is primarily sampled from the \emph{spike}-part of the prior will be
strongly shrunken towards zero if $v_0$ is sufficiently small, so that the posterior probability for $\gamma_j=v_0$ can be interpreted as the probability of exclusion of $\bm
\beta_j$ and the corresponding function $f_j$ from the model. A Beta prior for the mixture weights $w$ can be used to incorporate the analyst's prior knowledge about the sparsity of
$\bm \beta$ or, more practically, enforce sufficiently sparse solutions for overparameterized models.

We refer to the complete prior structure for $\alpha_j$ as a normal-mixture-of-inverse gamma (NMIG) distributions, denoted as $\alpha_j \sim \operatorname{NMIG}(v_0, w,
a_\tau,b_\tau)$. A similar NMIG prior has originally been suggested in \citet{Ishwaran:2005} for selecting single coefficients $\beta_j \sim N(0,v_j^2)$ in high-dimensional linear
models.

Integrating out $\tau_j^2$ and $\gamma_j$ from $\alpha_j \sim N(0, \nu_j^2=\gamma_j \tau_j^2)$, keeping $w$ fixed, the bimodal Inverse Gamma prior induces the mixture of two scaled
t-distributions $$\alpha_j|w \sim (1-w) t(df, s_0) + w\, t(df, s_1),$$ with $df=2a_\tau$, $s_0=\sqrt{v_0 b_\tau/a_\tau}$, $s_1=\sqrt{b_\tau/a_\tau}$, as a spike and slab prior
directly on $\alpha_j$. This may suggest to put other spike and slab priors directly on $\alpha_j$, in particular a bimodal mixture of normals
\begin{equation}
(1-w) N(0, \nu_{0j}^2) + w N(0, \nu_{1j}^2), \label{NormMix}
\end{equation} with $\nu_{1j}^2 \gg \nu_{0j}^2$, introduced by
\citet{George:McCulloch:1993} for selecting scalar coefficients in high dimensional linear models. Another alternative seems to be to replace $N(0, \nu_{0j}^2)$ by a point mass at
zero, but this works only for Gaussian regression models where it is possible to integrate out parameters analytically from full conditionals for indicator variables.

We prefer the NMIG prior for $\alpha_j$, inducing the student spike and slab prior, for several reasons: First, as noted in \citet{Ishwaran:2005} and supported by our own experience,
posterior inference and model selection is relatively robust with respect to hyperparameters in the NMIG hierarchy. It can be more difficult to specify $\nu_{0j}, \nu_{1j}$ for the
Gaussian spike and slab prior \eqref{NormMix}. Second, Section 5 of \citet{Ishwaran:2005} provides theoretical arguments in favor of a bimodal continuous prior for variances, as in
the NMIG and peNMIG prior, whereas \eqref{NormMix} is a bimodal discrete prior $(1-w) I(\nu_j^2=\nu_{0j}^2) + w I(\nu_j^2=\nu_{1j}^2)$. Furthermore, the favorable shrinkage
properties of the NMIG prior induce desirable shrinkage properties for the coefficient vector $\bm\beta_j$, see Subsection~\ref{sec:methods:shrinkage}.

Entries of the vector $\bm\xi_j$ are assigned the prior distribution
\[
\xi_{jk}|m_{jk} \sim N(m_{jk},1), \quad m_{jk} \sim \frac{1}{2}\delta_1(m_{jk}) + \frac{1}{2}\delta_{-1}(m_{jk}).
\]
i.e. we assume i.i.d. mixtures of two Gaussian distributions with expectation $\pm1$. Although the marginal prior for $\xi_{jk}$ still has zero expectation, the bivariate mixture
assigns most of the prior mass close to the multiplicative identity (with positive or negative sign). This enables the interpretation of $\alpha_j$ as the ``importance'' of
the $j$-th coefficient block and yields a marginal prior for $\bm\beta_j$ that is less concentrated on small absolute values than with a standard assumption like $\xi_{jk} \sim
N(0,1)$.

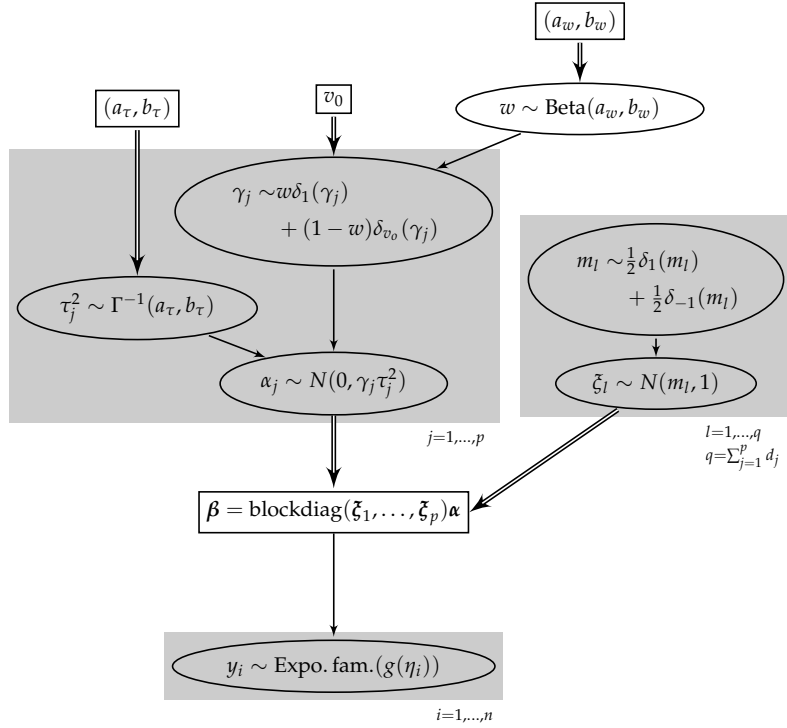
\begin{figure}[!tp]
\begin{scriptsize}
\begin{center}
peNMIG: NMIG with parameter expansion\\
$\phantom{bla}$\\
\begin{tikzpicture}
[hyper/.style={rectangle, draw, semithick},
 param/.style={ellipse, draw, semithick},
 parent/.style = {<-, shorten <=1pt, >=latex', semithick},
 child/.style = {->, shorten <=1pt, >=latex', semithick},
 determparent/.style = {<-, double, shorten <=1pt, >=latex', semithick},
 determchild/.style = {->, double, shorten <=1pt, >=latex', semithick}] 
    \node[param](alpha)  {$\alpha_j \sim N(0, \gamma_j\tau_j^2)$};
    \node[param](gamma)  [above =4em of alpha] {$\begin{aligned}\gamma_j \sim &  w  \delta_1(\gamma_j) \\& {} + (1-w)\delta_{v_o}(\gamma_j)\end{aligned}$}
        edge[child] (alpha);
    \node[hyper](v0) [above =2em of gamma]    {$v_0$}
        edge[determchild] (gamma);
    \node[param](w)  [above right=3em of gamma] {$w \sim \operatorname{Beta}(a_w, b_w)$}
        edge[child] (gamma);
    \node[hyper](hyperw)  [above =2em of w]    {$(a_w, b_w)$}
        edge[determchild] (w);
    \node[param](tau)  [above left=2em of alpha] {$\tau_j^2 \sim \Gamma^{-1}(a_\tau, b_\tau)$}
        edge[child] (alpha);
    \node[hyper](hypertau)   [above=7em of tau]  {$(a_\tau, b_\tau)$}
        edge[determchild] (tau);
     \begin{pgfonlayer}{background}
    \node[rectangle, fill=gray!40, fit = (alpha) (tau) (gamma) ,  label=-40:$^{j = 1,\dots,p}$]{};
    \end{pgfonlayer}
    \node[hyper](beta)  [below= of alpha]    {\mbox{$\bm \beta = \operatorname{blockdiag}(\bm
    \xi_1,\dots,\bm\xi_p) \bm \alpha$}}
        edge[determparent]    (alpha);
    \node[param](xi)    [right=5em of alpha]    {$\xi_l \sim N(m_l, 1)$}
     edge[determchild]    (beta.east);
    \node[param](m)     [above=1em of xi]       {$\begin{aligned} m_l
    \sim &
    \tfrac{1}{2}\delta_{1}(m_l)\\ & {} + \tfrac{1}{2}\delta_{-1}(m_l)\end{aligned}$} edge[child] (xi);
    \begin{pgfonlayer}{background}
    \node[rectangle, fill=gray!40, fit = (xi) (m), label=-90:$\phantom{lkjadghkljghlkjdhg}^{l =
    1,\dots,q}_{q=\sum_{j=1}^p d_j}$](fitxim){};
    \end{pgfonlayer}
     \node[param] [below=5em of beta] (y) {$y_i \sim
    \operatorname{Expo.fam.}(g(\eta_{i}))$} edge[parent] (beta);
    \begin{pgfonlayer}{background}
     \node[rectangle, fill=gray!40,  fit = (y), label=-20:$^{i =
     1,\dots,n}$](yfit){};
    \end{pgfonlayer}
\end{tikzpicture}
  \caption[DAG of peNMIG prior]{Directed acyclic graph of NMIG
  model with parameter expansion.\newline Ellipses are stochastic nodes, rectangles are deterministic/logical nodes.
  Single arrows are stochastic edges, double arrows are logical/deterministic
  edges.}
  \label{parExpDAG}
\end{center}
\end{scriptsize}
\end{figure}

The prior specification for $\bm\beta_j$ is completed by assuming prior independence between $\alpha_j$ and $\bm \xi_j$. We write $\bm\beta_j \sim \operatorname{peNMIG}(v_0, w,
a_\tau, b_\tau)$ for the complete prior structure also summarized as a directed acyclic graph in Figure~\ref{parExpDAG}.

The main advantage of the $\operatorname{peNMIG}$-prior is that the dimension of the coefficient vector associated with updating $\gamma_j$ and $\tau^2_j$ is equal to one in every
penalization group, since the Markov blankets of both $\gamma_j$ and $\tau_j$ only contain the scalar parameter $\alpha_j$ instead of the vector $\bm \beta_j$. This is crucial in
order to avoid mixing problems that would arise in a conventional NMIG prior without parameter expansion (see web appendix A).
The vector $\bm \xi=(\bm \xi_1\tr, \dots,\bm \xi_p)\tr$ is decomposed into subvectors $\bm \xi_j,\; j=1,\dots,p,$ associated with the different model terms and their respective
entries $\alpha_j$ in $\bm\alpha$. Note that $\eta_i$ typically also includes terms that are not under selection, such as known offsets, a global intercept or covariate effects that
are forced into the model. Their coefficients are associated with weakly informative flat Gaussian priors.

For Gaussian responses, we assume an $\operatorname{IG}(a_\sigma,b_\sigma)$ prior for the variance $\sigma^2$.

\subsection{Shrinkage Properties}\label{sec:methods:shrinkage}

This section describes regularization properties of marginal priors for regression coefficients, implied by the hierarchical prior structure described in the previous section and
visualized in Figure \ref{parExpDAG}. We analyze marginal priors because it is their shape - and less that of the conditional priors - that determines the shrinkage properties.

For comparison with other shrinkage priors recently suggested for pure variable selection, i.e. for selecting single scalar regression coefficients rather than blocks of
coefficients, we first consider \textit{univariate marginal priors}. Omitting indices $j$ and the dependence on hyperparameters $a_\tau$, $b_\tau$, $a_w$, $b_w$ and $v_0$, the
marginal prior for a scalar coefficient $\beta$ is obtained by integrating out all other random variables
 appearing in the prior hierarchy, i.e.
\begin{equation}\label{eq:margprior:scalar}
  p(\beta=\alpha\xi) = \int
p(\alpha|\gamma,\tau^2)p(\frac{\beta}{\alpha}|m)\frac{1}{|\alpha|}p(m)p(\tau^2)p(\gamma|w)p(w)d\alpha dm d\tau^2  d\gamma dw.
\end{equation}

Whereas the marginal prior for the original NMIG spike-and-slab prior of \citet{Ishwaran:2005} can be derived analytically as a mixture of scaled t-distributions, the above integral
has no known closed form and has to be computed numerically.
%
%
The marginal NMIG prior has a finite spike around zero, corresponding to the first component of the scaled t-mixture, and a slab corresponding to the second component. In comparison,
the marginal peNMIG prior has heavier tails and an infinite spike at zero. Its shape is very close to the horseshoe prior which has favorable theoretical properties. The peNMIG prior
also looks similar to the original spike-and-slab prior suggested by \citet{Mitchell:Beauchamp:1988}. The tails of the marginal peNMIG prior are also heavy enough to imply
redescending score functions which ensures Bayesian robustness of the resulting shrinkage estimators. The shape of the score function is similar to that of an $\mathcal{L}_q$-prior
with $q \rightarrow 0$ and is fairly robust with respect to hyperparameters, see
Figure 5 in \blind{\citet{Scheipl:2010}}.

In summary, the peNMIG prior for scalar $\beta$ combines an infinite spike at zero with heavy tails. This desirable combination is similar to other shrinkage priors, including the
horseshoe prior and also the normal-Jeffreys prior \citep{Bae:Mallick:2004}, for which both robustness for large values of $\beta$ and very efficient estimation of sparse coefficient
vectors have been shown \citep{Carvalho:2010, Polson:Scott:2010}.

\begin{figure}[!tbp]
\begin{center}
  \includegraphics[width=\textwidth]{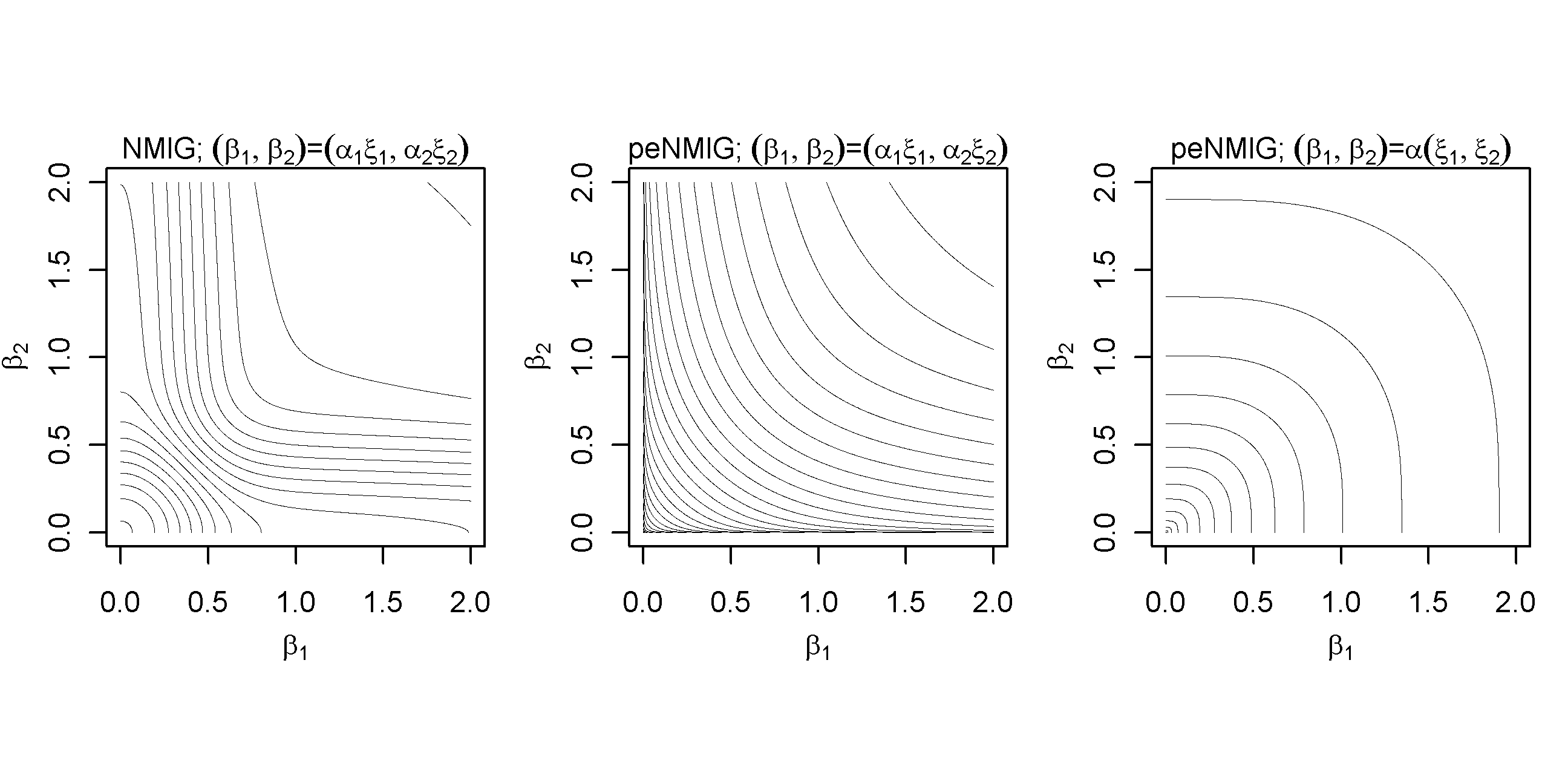}
  \caption[Constraint regions for $\beta$]{Contour plots of $\log
  p((\beta_1,\beta_2)\tr)$ for $a_\tau=5$, $b_\tau=50$, $v_0=0.005$, $a_w=b_w$
  for (from left to right) the NMIG prior for two coefficients from
  different penalization groups, the peNMIG prior for two coefficients from
  different penalization groups and the peNMIG prior for two coefficients from
  the same penalization group.} \label{fig:constraintRegion}
\end{center}
\end{figure}

A main advantage of our peNMIG prior is its generalization to \textit{multiple shrinkage} of blocks of coefficients vectors, both in terms of sampling and shrinkage properties. We
illustrate this for two-dimensional coefficients $(\beta_1,\beta_2)$, distinguishing two situations: First, $\beta_1$ and $\beta_2$ come from two different coefficient (or
penalization) groups, i.e. we have $\beta_1 = \alpha_1 \xi_1$ and $\beta_2 = \alpha_2 \xi_2$, with $\alpha_1$ independent from $\alpha_2$, in terms of the reparameterized
coefficients. Second, $\beta_1$ and $\beta_2$ are from the same block  of coefficients representing one penalization group, so that $(\beta_1,\beta_2) =\alpha (\xi_1,\xi_2)$, with
identical $\alpha_1 = \alpha_2 = \alpha$. Figure \ref{fig:constraintRegion} shows contour plots of $\log p(\beta_1,\beta_2)$ for these two situations and, for comparison, for the
original NMIG prior which applies only to the first situation.

The shape of the constraint region implied by the peNMIG prior (middle of Figure \ref{fig:constraintRegion}) has the convex shape of a $\mathcal{L}_q$-penalty function with $q<1$,
which has the desirable properties of simultaneous strong shrinkage of small coefficients and weak shrinkage of large coefficients due to its closeness to the $\mathcal{L}_0$
penalty.

The contours of the NMIG prior (left part of Figure \ref{fig:constraintRegion}) have different shapes depending on the distance from the origin. Close to the origin ($\beta< .3$),
they are circular and very closely spaced, implying strong ridge-type shrinkage: Coefficient values close to zero fall into the spike-part of the prior and will be strongly shrunk
towards zero. Moving away from the origin ($.3<\beta<.8$), the shape of the contours defining the constraint region morphs into a rhombus shape with rounded corners that is similar
to that produced by a Cauchy prior. Still further from the origin ($1<\beta<2$), the contours become convex and resemble those of the contours of an $\mathcal{L}_q$-penalty function
with $q<1$. Coefficient pairs in this region will be shrunk towards one of the axes, depending on which of their maximum likelihood estimators is bigger and their posterior
correlation. For larger values of coefficient pairs, the contours (not shown in Figure \ref{fig:constraintRegion}) imply ridge-type shrinkage.

The shape of the constraint region for coefficient pairs $(\beta_1,\beta_2) =\alpha (\xi_1,\xi_2)$ from the same penalization group (right part of Figure \ref{fig:constraintRegion})
resembles that of a square with rounded corners. Compared with the convex shape of the constraint region in the middle, this shape induces less shrinkage towards the axes and more
towards the origin or along the bisecting angle.

\subsection{Markov Chain Monte Carlo Algorithm}\label{sec:methods:MCMC}

Posterior inference and function selection is based on a blockwise Metropolis-within-Gibbs sampler. The sampler cyclically updates the nodes in Figure \ref{parExpDAG}. For Gaussian
responses it reduces to a Gibbs sampler. The full conditionals of the parameters $w$, $\tau_j^2$, $\gamma_j$, $j=1,\dots,p$, and the means $\bm{m} = (m_1,\dots,m_l,\dots,m_q)\tr$ of
the conditionally Gaussian variables $\xi|m_l \sim N\left(m_l, 1 \right)$, $m_l=\pm1$, are available in closed form and are included in Algorithm \ref{MCMCalgFCD} in the appendix.
They do not depend on the specific exponential family chosen for the responses.

The full conditionals for $\bm{\alpha} = (\alpha_1,\dots,\alpha_p)\tr$ and $\bm{\xi} = (\xi_1\tr,\dots,\xi_p\tr)\tr$ depend on the ``conditional'' design matrices $\bm X_\alpha = \bm
X \operatorname{blockdiag}(\bm\xi_1,\dots,\bm\xi_p)$ and \newline $\bm X_\xi = \bm X \operatorname{diag}(\operatorname{blockdiag}({\bm 1_d}_1,\dots,{\bm 1_d}_p)\bm\alpha)$,
respectively, where $\bm 1_d$ is a $d \times 1$ vector of ones and $\bm X = (\bm X_1,\dots,\bm X_p)$ is the concatenated design matrix. For Gaussian responses, the full conditionals
are given by
\begin{align}
\begin{split}
 \bm\alpha|\cdot &\sim N(\bm \mu_{\alpha}, \Sigma_{\alpha}) \text{ with}\\
 \bm \Sigma_{\alpha} &= \left(\frac{1}{\sigma^2}\bm X_\alpha^T\bm X_\alpha +
\operatorname{diag}\left(\bm \gamma \bm \tau^2\right)^{-1}\right)^{-1}, \; \bm\mu_j =\frac{1}{\sigma^2}\bm\Sigma_\alpha\bm X_\alpha^T\bm y,
\text{ and}\\
 \bm\xi|\cdot &\sim N(\bm \mu_{\xi}, \bm \Sigma_{\xi}) \text{ with}\\
 \bm \Sigma_{\xi} &= \left(\frac{1}{\sigma^2}\bm X_\xi^T\bm X_\xi +
\bm I \right)^{-1}; \; \bm\mu_j =\bm\Sigma_\xi\left(\frac{1}{\sigma^2}\bm X_\xi^T\bm y + \bm m \right).
\end{split}\label{F:fcdCoef}
\end{align}

For non-Gaussian responses, we use an MH algorithm with a penalized IWLS proposal (P-IWLS) based on an approximation of the current posterior mode, described in detail in
\citet{Brezger:Lang:2006} (Sampling scheme 1, Section 3.1.1). The MH step uses a Gaussian (i.e.\ second order Taylor) approximation around the approximate mode of the full
conditional as its proposal distribution. To decrease computational complexity, we modify the IWLS algorithm by using the mean of the proposal distribution of the previous step
instead of the posterior mode. Because of the prohibitive computational cost for large $q$ and $p$ (and low acceptance rates for non-Gaussian response for high-dimensional IWLS
proposals), neither $\bm \alpha$ nor $\bm \xi$ are updated all at once. Rather, both $\bm \alpha$ and $\bm \xi$ are split into $b_\alpha$ ($b_\xi$) update blocks that are updated
sequentially conditional on the states of all other parameters. For efficient and numerically stable draws from the multivariate Gaussian densities in \eqref{F:fcdCoef} or in the
IWLS algorithm, we use QR decompositions of the covariance matrices. Alternatively, Cholesky decompositions as in \citet{Lang:Brezger:2004} or \citet{Brezger:Lang:2006} may be
employed.

After updating the entire $\bm \alpha$- and $\bm \xi$-vectors, each subvector $\bm \xi_j$ is rescaled so that $|\bm \xi_j|$ has mean 1, and the associated $\alpha_j$ is rescaled
accordingly so that $\bm \beta_j = \alpha_j \bm \xi_j$ is unchanged:
\begin{align*}
\bm \xi_j &\rightarrow \frac{d_j}{\sum_i^{d_j} |\xi_{ji}|}\bm \xi_j \quad\text{and}\quad \alpha_j \rightarrow \frac{\sum_i^{d_j} |\xi_{ji}|}{d_j}\alpha_j.
\end{align*}
This rescaling is advantageous since $\alpha_j$ and $\bm \xi_j$ are not identifiable and thus their sampling paths can wander off into extreme regions of the parameter space without
affecting the fit, e.g. $\alpha_j$ becoming extremely large while entries in $\bm \xi_j$ simultaneously become extremely small. By rescaling, we retain the interpretation of
$\alpha_j$ as a scaling factor representing the importance of the model term associated with it and avoid numerical problems that can occur for extreme parameter values.

By default, starting values $\bm \beta^{(0)}$ are drawn randomly in three steps: First, we do 5 Fisher scoring steps with fixed, large hypervariances. Second, for each chain run in
parallel, Gaussian noise is added to this vector, and third its constituting $p$ subvectors are scaled with variance parameters $\gamma_j\tau_j^2\;( j=1,\dots,p)$ drawn from their
priors. This means some of the $p$ model terms are set close to zero initially, and the remainder is in the vicinity of their respective ridge-penalized maximum likelihood estimates.
Starting values for $\bm\alpha^{(0)}$ and $\bm\xi^{(0)}$ are then computed via $ \alpha^{(0)}_j = d_j^{-1} \sum_i^{d_j} |\beta^{(0)}_{ji}|$ and $\bm \xi^{(0)}_j = \bm\beta^{(0)}_j /
\alpha^{(0)}_j$.

\textit{Function selection}, i.e. selection of coefficient blocks $j=1,\dots,p$ can be based on the posterior inclusion probabilities $P(\gamma_j=1 | \bm y)$. Instead of estimating
them simply through the proportion of MCMC samples $(t)$ for which $\gamma_j^{(t)}=1$, which may have high sampling variance, we improve precision through the Rao-Blackwellized
estimate $\hat P(\gamma_j = 1| \bm y) = T^{-1} \sum_t P(\gamma_j^{(t)}=1|\cdot)$. The full conditional probabilities $P(\gamma_j^{(t)}=1|\cdot)$ are directly available from step 14
of the MCMC Algorithm \ref{MCMCalgFCD} (see appendix). Based on these quantities, we can evaluate the evidence for the effect of a continuous covariate being linear or nonlinear,
whether a model term has a relevant effect at all, which interaction terms are relevant, and so on.

\section{EMPIRICAL EVALUATION}\label{sec:eval}

\subsection{Simulations}\label{sec:sim}

To evaluate the performance of the peNMIG prior for function selection in generalized additive models, we conducted extensive simulations representing scenarios of different
complexity comprising different response types, sample sizes, signal to noise ratios, correlated and uncorrelated covariates, varying degrees of concurvity, and predictors of either
high or low sparsity. As competitors, we considered component-wise boosting \citep{mboost}, ACOSSO \citep{Storlie:Bondell:Reich:2010}, as well as SPAM \citep{Ravikumar:2009}, the
double shrinkage GAM proposed by \citet{Marra:Wood:2011}, the HGAM approach of \citet{Meier:Geer:Buehlmann:2009} and finally Bayesian additive regression trees
\citep{Chipman:George:McCulloch:2010} as a non-parametric, ``black-box prediction'' alternative. As benchmark, we used a conventional GAM based on the true model structure. Note that
ACOSSO, SPAM and HGAM are available for Gaussian responses only. Predictive deviance and the ability to recover the correct model complexity were used as performance measures.
Detailed description and graphical summaries of these simulations can be found in Section \ref{sec:simApp} of the appendix. Extensive additional simulation studies are described in
the first author's dissertation \blind{\citep{Scheipl:2011}}.

The main conclusion that can be drawn from the simulations is that the proposed approach is highly competitive to previous suggestions while having the advantage of being applicable
both for generalized exponential family regression (while most previous suggestions are restricted to Gaussian responses) and a much broader class of model terms (most previous
suggestions are restricted to univariate smooth functions). Estimation based on the peNMIG prior typically achieves good estimation performance simultaneously with high accuracy in
determining the correct model specification. It is robust against correlated covariates and low signal-to-noise ratios. Prediction accuracy is robust against concurvity, however very
strong concurvity of course leads to diminished selection accuracy. Selection of large coefficient blocks such as random effects for non-Gaussian response can be problematic: For
both Poisson and binary responses, the selection accuracy in this case was very low, albeit without adverse effects on the estimation of the coefficients.

Robustness to hyperparameter settings was investigated by running the simulations for combinations of $v_0 = 0.01, 0.005, 0.00025$ and $(a_\tau, b_\tau)=(5, 25), (5, 50), (10, 30)$.
Prediction accuracy was very robust against different hyperparameter configurations in all the settings we considered while variable selection and model choice were more sensitive to
varying hyperparameters, because estimated posterior inclusion probabilities for small effects are sensitive to the value for $v_0$. This parameter controls the threshold of
relevance of the model terms: in general, very small $v_0$ means small effects are more likely to be included in the model, while larger $v_0$ yield more conservative selection
properties. However, the conducted simulations and the applications presented in the following sections provide a solid foundation for the choice of appropriate values for a wide
range of applied problems. We have successfully fitted all of the application examples and the binary classification data discussed in the following section with a default prior that
uses $v_0=0.00025, (a_\tau, b_\tau)=(5, 25)$ and $(a_w, b_w)=(1,1)$.

\subsection{Binary Classification Benchmarks}\label{sec:app:uci}

We use a collection of 21
data sets for binary classification
to investigate the performance of our
approach on some well known benchmarks. The same collection has previously been
used for benchmarking in \citet{Meyer:Hornik:Leisch:2003} which contains
some more details on the datasets that we use.
\begin{table}[ht]
\begin{center}
\begin{tabular}{l|rrrr}
  data set & $n$ & covariates & of which factors & balance\\
  \hline
  \data{BreastCancer} & 683 & 9 & 9 & 0.54 \\
  \data{Cards} & 653 & 15 & 5 & 0.83 \\
  \data{Circle} & 1200 & 2 & 0 & 0.97 \\
  \data{Heart1} & 296 & 13 & 5 & 0.85 \\
  \data{HouseVotes84} & 232 & 16 & 0 & 0.87 \\
  \data{Ionosphere} & 351 & 33 & 0 & 0.56 \\
  \data{PimaDiab} & 768 & 8 & 0 & 0.54 \\
  \data{Sonar} & 208 & 60 & 0 & 0.87 \\
  \data{Spirals} & 1200 & 2 & 0 & 1.00 \\
  \data{chess} & 3196 & 36 & 1 & 0.91 \\
  \data{credit} & 1000 & 24 & 9 & 0.43 \\
  \data{hepatitis} & 80 & 19 & 0 & 0.19 \\
  \data{liver} & 345 & 6 & 0 & 0.72 \\
  \data{monks3} & 554 & 6 & 4 & 0.92 \\
  \data{musk} & 476 & 166 & 0 & 0.77 \\
  \data{promotergene} & 106 & 57 & 57 & 1.00 \\
  \data{ringnorm} & 1200 & 20 & 0 & 1.00 \\
  \data{threenorm} & 1200 & 20 & 0 & 1.00 \\
  \data{tictactoe} & 958 & 9 & 9 & 0.53 \\
  \data{titanic} & 2201 & 3 & 1 & 0.48 \\
  \data{twonorm} & 1200 & 20 & 0 & 1.00 \\
   \hline
\end{tabular}
 \caption[Characteristics of UCI data sets]{Characteristics of UCI data sets.
 ``Balance'' is the ratio between the number of observations in the larger class
 and the number of observations in the smaller class, i.e. it is $1$ if the data
 set is balanced.}
  \label{uciInfo}
\end{center}
\end{table}
Table \ref{uciInfo}
gives an overview of the datasets and their characteristics.
We evaluate prediction performance based on the deviance values for
a 20-fold cross validation on each dataset. Predictive deviance
$\bar D$ is defined as twice the average negative log
likelihood $\bar D = -2/n_P \sum^{n_P}_{i=1} L(y_{P,i}, \hat\eta_{P,i})$ in the
test sample where $\bm{y}_P$ and $\bm{\hat\eta}_P$ are the out-of-sample
responses and the posterior mean of the linear predictor for the test
sample. The size of the test sample is $n_P$.
As for
the experiments with simulated data, we use component-wise boosting
with separate base learners for the linear and smooth parts of covariate
influence and compare prediction performance of the boosting
models to our approach. For boosting, we determine the stopping
iteration $m_{\operatorname{stop}}$ based on the out-of-bag risk in
10 bootstrap samples of the training sample.

The second metric we are interested in is the parsimony of the estimated models. We simply count the number of model terms (or
baselearners for boosting) included in the model, i.e., we count the model terms with marginal posterior inclusion
probabilities greater than $0.5$.  For boosting, we count a baselearner as included if it was selected at least once before
iteration $m_{\operatorname{stop}}$ in more than half of the bootstrap samples used to determine $m_{\operatorname{stop}}$.

The data are preprocessed in order to preempt numerical problems: All covariates with less than six unique values are treated
as factor variables. All numeric covariates are logarithmized if their absolute skewness is larger than two and standardized to
have mean zero and unit standard deviation. Incomplete observations are removed. All numeric covariates are associated with
both a linear and a smooth effect. For data sets \data{credit, Cards, Heart1, Ionosphere, hepatitis, Sonar} and \data{musk}, we
use NMIG instead of peNMIG for terms with $d=1$ (i.e. linear terms and binary factors) to reduce the posterior's
dimensionality.
Estimates are based on samples from eight parallel chains with a burn-in of 1000 iterations, followed by a sampling phase of
5000 iterations of which we save every fifth. We report results for $(a_\tau, b_\tau)=(5,25)$, $w \sim
\operatorname{Beta}(1,1)$ and $v_0 \in \{0.005,  0.00025\}$. Results with $(a_\tau, b_\tau)=(5, 50)$ and $w \sim
\operatorname{Beta}(20, 40)$ were qualitatively similar and are omitted for clarity of presentation. An unabridged description
is in \blind{\citet[][Ch.~4.2]{Scheipl:2011}}.

\begin{figure}[!tbp]
\begin{center}
\includegraphics[width=\textwidth]{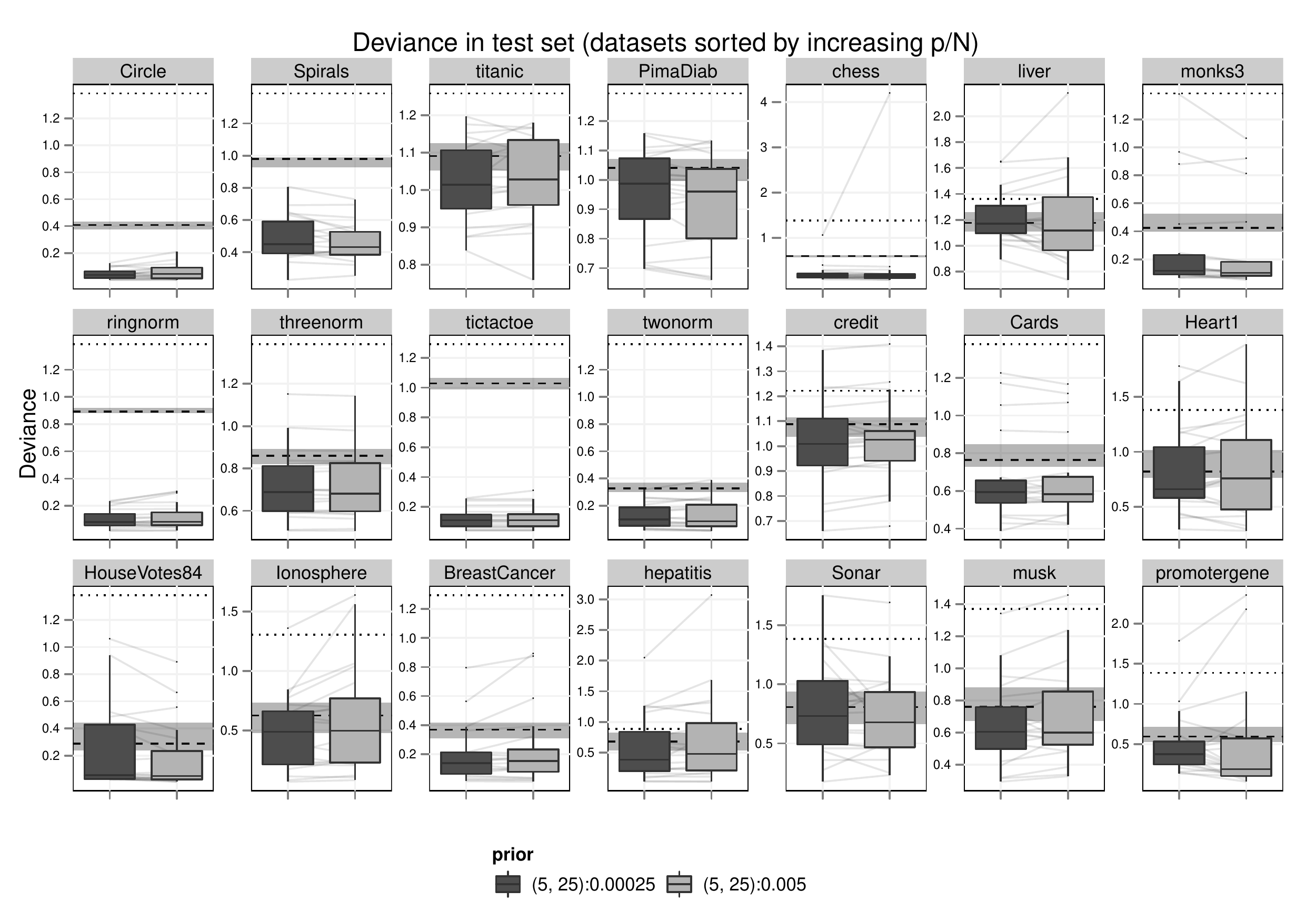}
  \caption[UCI data: predictive deviance]{UCI data I: Predictive Deviances for
  20-fold crossvalidation. Boxplots show results for the different prior settings,
   the horizontal ribbon shows results for \package{mboost}:
   shaded region gives IQR, dashed line represents median. Dark grey lines
   connect results for the same fold. Dotted line gives predictive deviance of
   the null model on the full data set.}
  \label{uci_predDev}
\end{center}
\end{figure}
Figure \ref{uci_predDev} shows the prediction accuracy for the 21
datasets. Most outliers with large deviances are due to the sampler getting stuck for some
of the parallel chains in specific folds for some of the data sets. By rerunning the analysis
with different starting values or random seeds or manual postprocessing of the
posterior samples, these presumably could have been avoided. We include them
unchanged to provide a more realistic picture of the reliability of our
approach. Practicioners should always check acceptance rates and convergence
diagnostics when using MCMC-based methods.
Note that the predictive performance of our approach is usually more variable
than that achieved by \package{mboost} but has lower median predictive deviances in all
of the datasets for all prior specifications (Results for $(a_\tau, b_\tau)=(5,
50)$ and $w \sim \operatorname{Beta}(20, 40)$ not shown). Predictive
performance is very robust against different hyperparameter settings,
even for large $p/n$ ratio where the influence of the hyperparameters on the
posterior is relatively stronger.
\begin{figure}[!tbp]
\begin{center}
  \includegraphics[width=\textwidth]{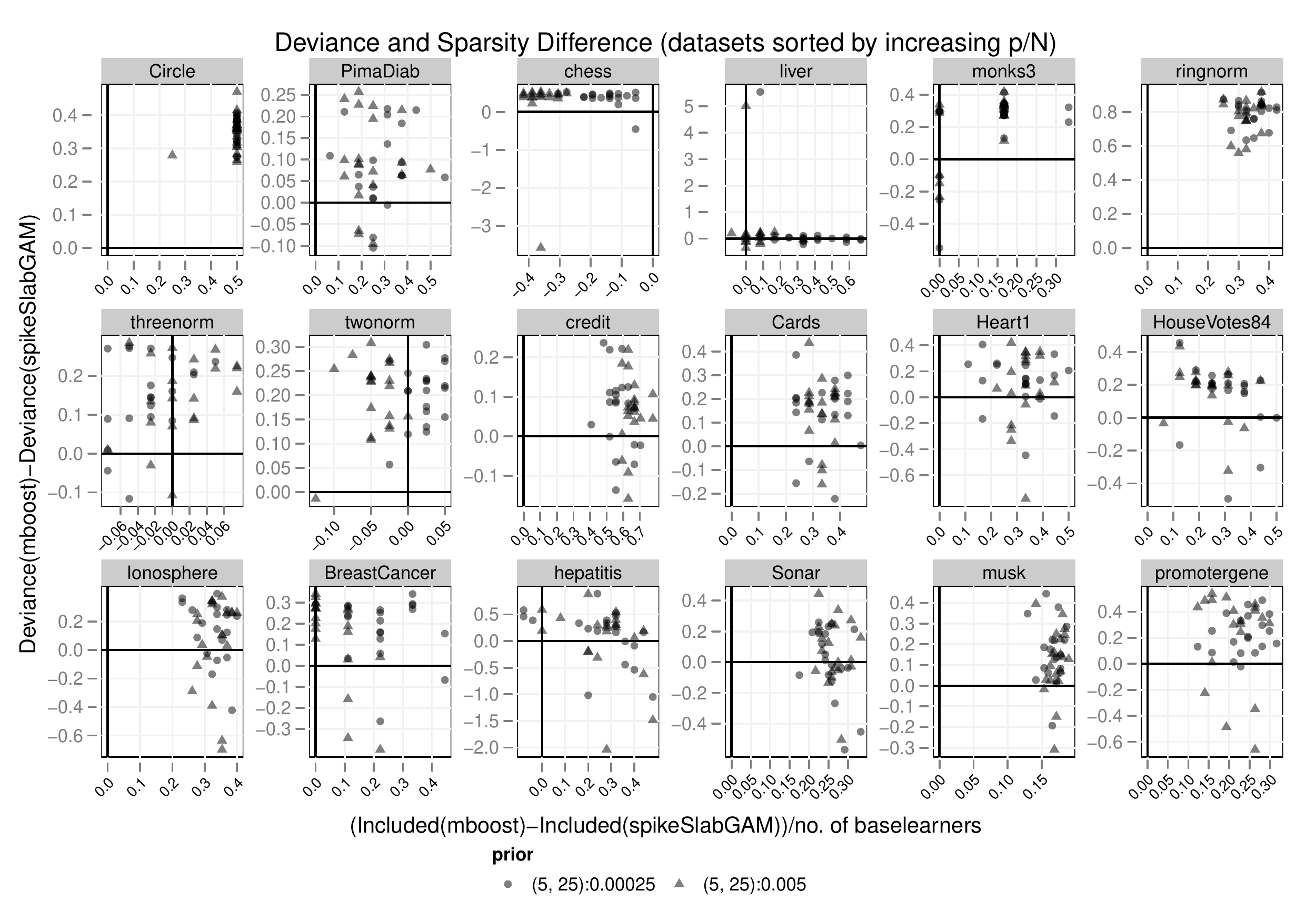}
  \caption[UCI data: sparsity vs. predictive deviance]{UCI data I: Difference
  in proportion of included model terms versus differences in predictive deviances.
  Positive values denote smaller deviances / models for our approach compared to \package{mboost}.
  Results for \data{Spirals, titanic} and \data{tictactoe} not shown because there were no differences in sparsity.}
  \label{uci_sparsityPredDev}
\end{center}
\end{figure}
To investigate the parsimony of the fitted models, i.e. whether
equivalent or better prediction can be achieved by simpler models,
we plot the differences in predictive deviances versus the
difference in the proportion of potential model terms included in
the models in Figure \ref{uci_sparsityPredDev} (Results for
\data{Spirals, tictactoe} and \data{titanic} not shown because there
were no differences in sparsity). Positive values on the
vertical axis indicate smaller deviance for our approach, and positive
values on the horizontal axis indicate a sparser fit for our
approach. Figure \ref{uci_sparsityPredDev} shows that
our approach achieves its relatively more precise predictions with smaller
models on the large majority of the benchmark data sets. The only exceptions are datasets
\data{chess}, where the increased precision is achieved at the cost of less sparse models,
and, to a much lesser extent, \data{twonorm} and \data{threenorm}.
Neither absolute performance nor performance relative to
boosting seems to be tied to any of the easily observable
characteristics of the data sets (i.e.~$p$, $n$, $p/n$, or
balancedness). No clear picture emerges for the differences between the
prior specifications: As expected, a smaller value for $v_0$
tends to yield larger models, cf.~datasets \data{chess, twonorm,
Ionosphere, musk}, but there are counterexamples as well,
e.g.~\data{threenorm}. Both predictive deviance and sparsity results
are more sensitive towards $v_0$ than towards $(a_\tau,b_\tau)$ (Results for
$(a_\tau, b_\tau)=(5, 50)$ not shown).
Using an informative prior $w \sim \operatorname{Beta}(20, 40)$ to
enforce model sparsity has no appreciable effect and does
not influence prediction quality (Results not shown).

More generally, the performance of our approach shows that it is very competitive to componentwise boosting and that neither
relative nor absolute performance deteriorate in very high-dimensional problems (cf.~results for \data{musk} with $n=476$ and
$332$ potential model terms, of which $166$ are smooth terms.).

\subsection{Case Study: Survival of Surgical Patients with Severe
Sepsis}\label{sec:app:intensive}

\paragraph{Data:}
We use data on the survival of 462 patients with severe sepsis that
was collected in the intensive care unit of the Department of
Surgery at Munich's Gro{\ss}hadern hospital between March 1, 1993,
and February 28, 2005. \citet{Hofner:2010} have previously analysed
this data set. The follow-up period was 90 days after the beginning
of intensive care, with one drop-out after 66 days and 179 patients
surviving the observation period.

\paragraph{Models:}
We use a piecewise exponential model (PEM)
\citep[Ch.~9]{Fahr:Tutz:2001} to model the hazard rate
$\lambda(t,\bm x)$ of the underlying disease process, i.e. for fixed
time intervals defined by cutpoints $\bm\kappa=(\kappa_0=0,\kappa_1,
\dots,\kappa_I=t_{\max})$, where $t_{\max}$ is the maximal follow-up
time, the hazard rate for subject $i$ at time $t, \;\kappa_{j-1} < t
\leq \kappa_{j}$ in the $j^{th}$ interval is given by
\[
 \lambda(t, \bm x_i) = \exp\left(g_0(j) + \sum_{l=1}^Lg_l(j)v_{il}(j) + \sum_{m=1}^Mf_m(u_{im}(j)) + \bm z_i(j)'\bm\gamma\right)
\]
where $g_0(j)$ represents the baseline hazard rate in interval $j$, $g_l(j)$, $l=1,\ldots,L,$ are time-varying effects of
covariates $v_{il}(j)$, $f_m(u_{im}(j))$, $j=1,\ldots,J,$ are nonlinear effects of continuous covariates $u_{im}(j)$ and $\bm
z_i(j)'\bm\gamma$ contains linear, parametric effects. All time-dependent quantities are assumed to be piecewise constant on
the intervals such that e.g. $g_0(t)=g_0(j)$ for all $t\in(\kappa_{j-1},\kappa_j]$. The interval borders $\bm \kappa = (0, 5,
15, 25, \ldots, 85, 90)$ were chosen based on the shape of a nonparametric estimate of the marginal hazard rate. The likelihood
for the PEM is equivalent to that of a Poisson model with (1) one observation for each interval for each subject, yielding 2826
pseudo-observations in total, (2) offsets $o_{ij}=\max(0, \min(\kappa_{j}-\kappa_{j-1}, t_i - \kappa_{j-1}))$, where $t_i$ is
the observed time under risk for subject $i$ and (3) responses $y_{ij}$ equal to the event indicators $\delta_{ij}$, with
$\delta_{ij}=0$ if subject $i$ survived interval $j$ and $\delta_{ij}=1$ if not.

Our aim is twofold: We want to (1) estimate a model that allows
assessment of the influence of each available covariate on the
prognosis of patients, accounting for possibly time-varying and/or
nonlinear effects and (2) use this setting to evaluate the stability
of the selection and estimation of increasingly complex models on
real data. Important covariates included in the analysis are for
example the age of the patient, the haemoglobin concentration, the
presence of a fungal infection, different types of operations and
the Apache II score (a measure for the severity of disease), see
\citet{Hofner:2010} for a complete description.

\paragraph{Full Data Results}
\begin{figure}[!tbp] \centering
\includegraphics[width=\textwidth]{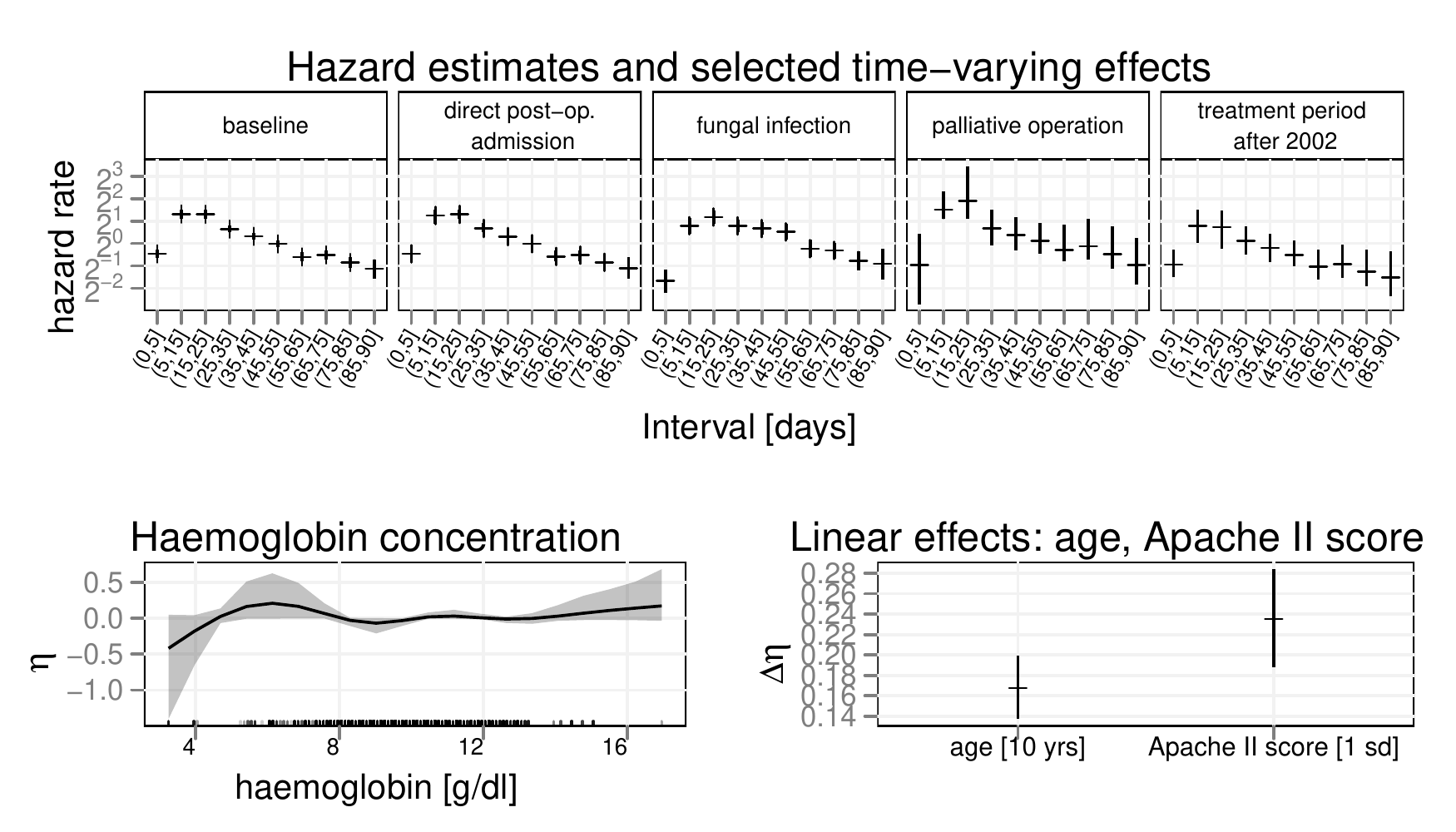}
\caption{Posterior means of effects with (pointwise) 80\% credible
intervals. Top: Baseline hazard rate and baseline hazard rate plus
the time-varying and time-constant effects for direct postoperative
admission, presence of a fungal infection, palliative operation and
beginning of treatment after 2002. Bottom: smooth effect of
haemoglobin concentration and linear effects of age (10 year
increase) and Apache II score, a measure for disease severity
(increase of score by 1 standard deviation).}
\label{fig:intensiveEffects_AllTV}
\end{figure}
\begin{table}[ht]
\begin{small}
\begin{center}
\begin{tabular}{l|c}
 Term & $P(\gamma=1)$ \\
  \hline
  MRF(Interval) & 1.00 \\
  palliative operation & 0.19 \\
  treatment period & 0.71 \\
  Age, linear & 0.99 \\
  Apache II score, linear & 1.00 \\
  Haemoglobin concentration, smooth & 0.38 \\
  MRF(Interval):direct postoperative admission & 0.28 \\
  MRF(Interval):fungal infection & 1.00 \\
  MRF(Interval):palliative operation & 0.38 \\
  MRF(Interval):treatment period & 0.13 \\
   \hline
\end{tabular}
\end{center}
\end{small}
\caption{Posterior means of marginal inclusion probabilities
$P(\gamma=1)$ (only given for terms with $P(\gamma=1)> .1$).}
\label{tab:intensiveEffects_AllTV}
\end{table}
We perform term selection for a maximal model which includes the (linear and non-linear) effects of all 20 covariates as well as their time-varying effects, i.e. 48 potential model
terms with 262 coefficients in total. Hyperparameters were set to the default values determined in the simulation studies, i.e. $a_\tau=5, b_\tau=25, v_0=0.00025$ and $a_w=b_w=1$.
Estimates are based on 8 parallel chains running for 20000 iterations each after a burn-in of 500 iterations, with every 10$^{th}$ iteration saved. We use a first order random walk
prior for the log-baseline and the time-varying effects to regularize their roughness, i.e.,~we use an intrinsic GMRF prior (on the line) for the piecewise constant time-varying
quantities (denoted as MRF(Interval) in the following).

The estimated marginal inclusion probabilities indicate a fairly
sparse model, with posterior marginal inclusion probabilities
greater than $0.10$ for only 10 terms, as shown in Table
\ref{tab:intensiveEffects_AllTV}. The estimated effects for this
subset of terms are visualized in Figure
\ref{fig:intensiveEffects_AllTV}.
\begin{figure}[!tbp] \centering
\includegraphics[width=\textwidth]{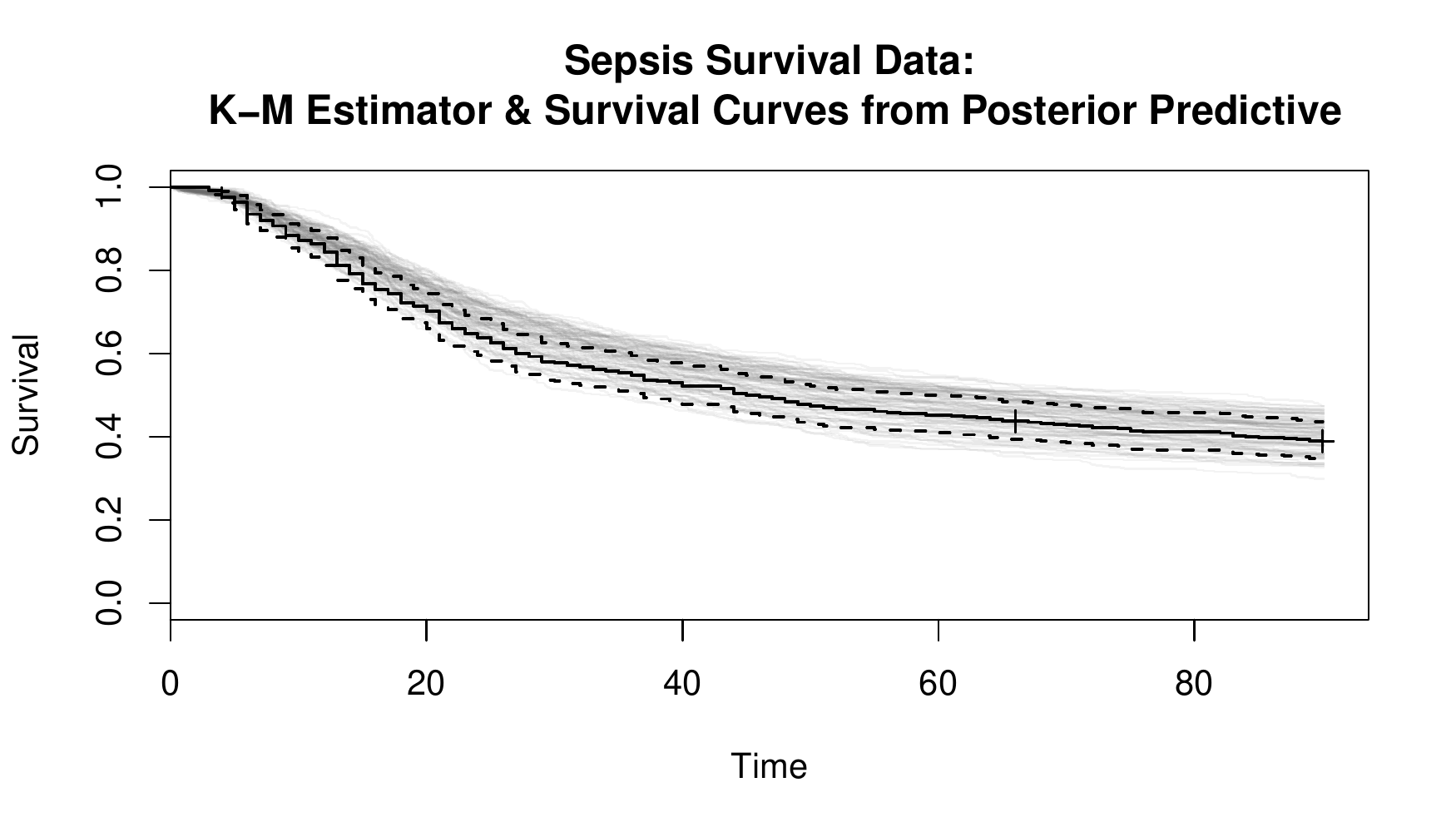}
\caption{Kaplan-Meier estimate of the survival curve for observed
data in black. Grey overlays are survival curves for 100 replicates
of survival time vectors generated from the posterior predictive
distribution.} \label{fig:intensivePostPredCheck_AllTV}
\end{figure}
To verify the suitability of the model, we perform a posterior
predictive check and generate 100 replicates of survival times from
the posterior predictive distribution. Figure
\ref{fig:intensivePostPredCheck_AllTV} indicates that the fit is
satisfactory, although there seems to be a tendency to overestimate
survival rates until about day 70.

\paragraph{Predictive Performance Comparison}
We subsample the data 20 times to construct independent training
data sets with 415 patients each and test data sets with the
remaining 47 patients to evaluate the precision of the resulting
predictions and compare predictive performance to that of equivalent
component-wise boosting models fitted with \package{mboost}. Results
for our approach are based on 8 parallel chains, each running for
5000 iterations after 500 iterations of burn-in, with every fifth
iteration saved. Component-wise boosting results are based on a
stopping parameter determined by a 25-fold bootstrap of the training
data, with a maximal iteration number of 1500.

The previous analysis by \citet{Hofner:2010} has used expert
knowledge to define a set of six covariates forced into the model
(indicators for presence of malignant primary disease, palliative
operation and beginning of treatment after 2002, as well as sex, age
and Apache II score). We compare results for four model
specifications of increasing complexity that suggest themselves: a
model with only the main effects of the pre-selected covariate set,
a model with main effects and time-varying effects for the
pre-selected covariate set, a model with main effects for all 20
covariates and the model with main effects and time-varying effects
for all 20 covariates which was applied to the whole data set (see
above). As in the previous section, main effects for numerical
covariates such as age were split into linear and non-linear parts.
\begin{figure}[!tbp] \centering
\includegraphics[width=\textwidth]{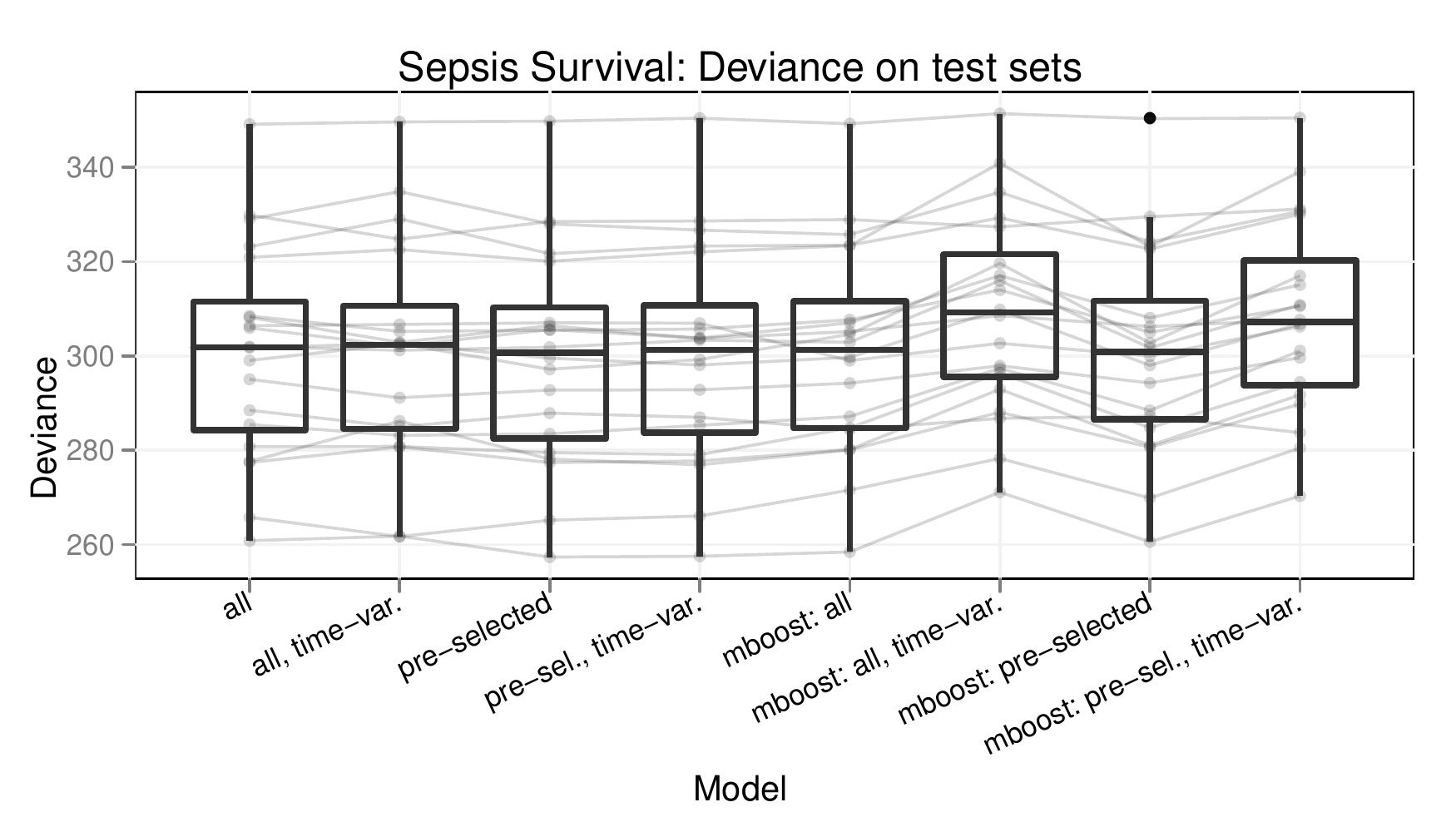}
\caption{Predictive deviances for 20 subsampling test sets for the
sepsis survival data (lower is better). Grey lines connect results
from identical folds.} \label{fig:intensive_CVPredDev}
\end{figure}
Figure \ref{fig:intensive_CVPredDev} shows the predictive deviances
achieved by the different model specifications. Predictive deviance
is defined as $-2 \sum^{N_t}_{i}\sum^{J(i)}_{j}
\delta_{ij}(\log(\hat\lambda_j) + \hat\eta_{ij}) -
o_{ij}\hat\lambda_j\exp(\hat\eta_{ij})$, where $i=1,\ldots,N_t$
indicates the subjects in the test set and $j=1,\ldots,J(i)$
indicates the intervals in which individual $i$ was under risk,
$\hat\lambda_j$ and $\hat\eta_{ij}$ are the respective posterior
predictive means. For this data set, models with higher maximal
complexity seem to offer no relevant improvement in terms of
prediction accuracy compared to the simplest model based only on the
pre-selected covariate set without time-varying effects. Most of the
models yield essentially equivalent predictions. However, it is
reassuring to see that the predictive performance of our approach is
not degraded at all by the specification of vastly over-complex
models in a setting where the underlying structure seems to be
fairly simple. In contrast, prediction accuracy for component-wise
boosting decreases markedly for the models including time-varying
effects in this setting.

The stability of the marginal term inclusion probabilities across
subsamples is fairly good, indicating that the term selection is
robust to small changes in the data. All model specifications
identified essentially the same subset of important effects from the
set of pre-selected covariates (i.e., indicators for palliative
operation and beginning of treatment after 2002 and linear effects
of age and Apache II score), and also the same time-varying effects
(i.e., time varying effects for palliative operation and beginning
of treatment after 2002).
\begin{figure}[!tbp] \centering
\includegraphics[height=.9\textheight]{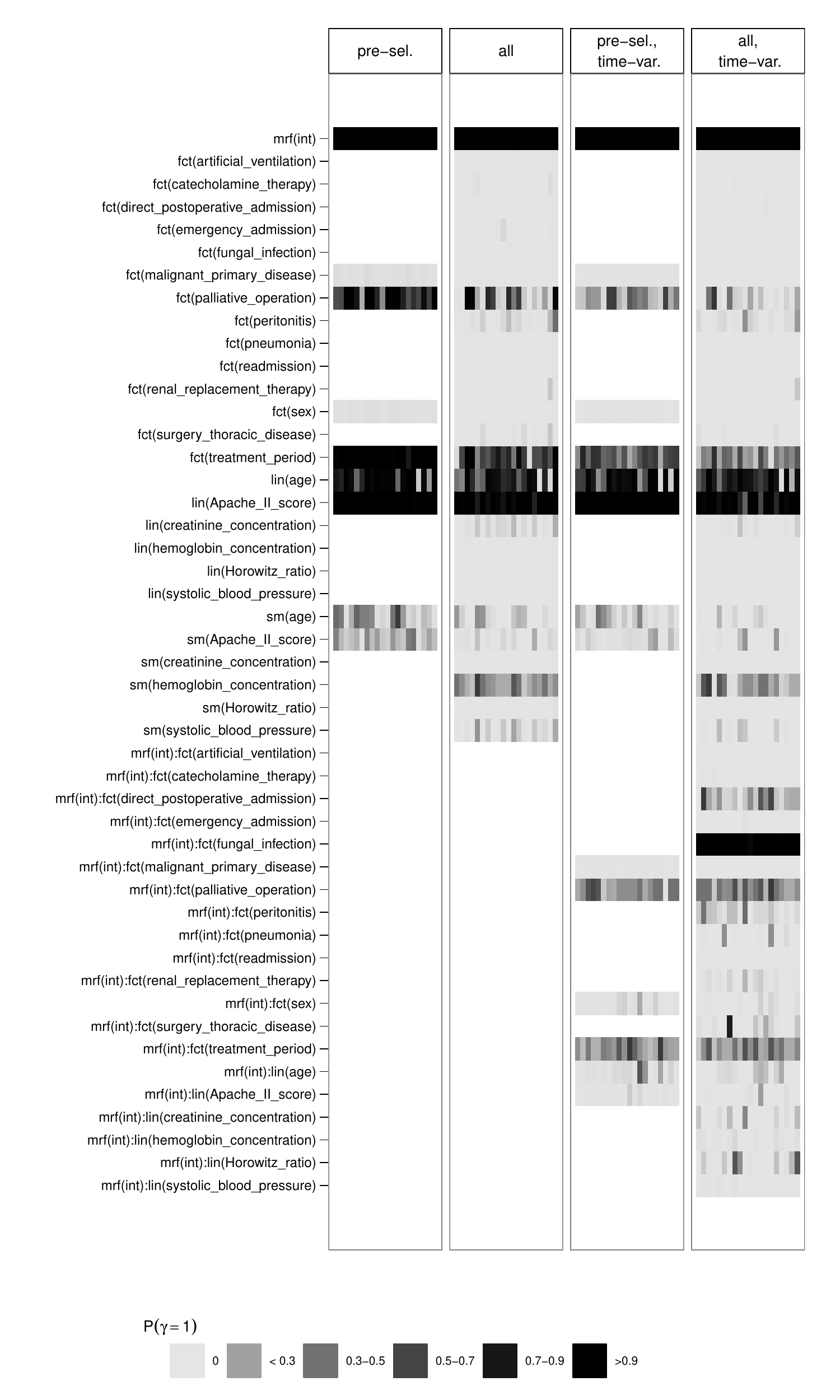}
\caption{Posterior means of inclusion probabilities $P(\gamma=1)$
across 20 subsampled training data sets for the 4 model
specifications.} \label{fig:intensive_CVIncProb}
\end{figure}
Figure \ref{fig:intensive_CVIncProb} shows the posterior means of
inclusion probabilities $P(\gamma=1)$ across 20 subsampled training
data sets for each of the 4 model specifications.

Additional case studies for geoadditive regression of net rent levels in Munich and an additive mixed model for binary responses from a large study on hymenoptera venom allergy can
be found in Section \ref{sec:CaseStudiesApp} of the appendix.

\section{CONCLUSIONS}\label{sec:conclusions}

In this paper, we have proposed a general Bayesian framework for conducting function selection in exponential family structured additive regression models. Inspired by stochastic
search variable selection approaches and the general idea of spike-and-slab priors, we introduced a non-identifiable multiplicative parameter expansion where selection or deselection
of coefficient batches (such as parameters representing a spline basis or random intercepts) is associated with a scalar scaling factor only. This reparameterization alleviates the
notorious mixing problems that would appear in a naive implementation of our prior structure.

The main advantages of the proposed peNMIG prior structure are (1) its general applicability for various types of responses (in particular non-Gaussian responses), (2) the
availability of a supplementing R package that makes all methods immediately accessible and reproducible, and (3) the good performance demonstrated in simulations and applications
with fairly low sensitivity with respect to hyperparameter settings and substantiated in theoretical investigations of the shrinkage properties of peNMIG. The class of models that
can be fitted in this framework can be extended fairly easily by considering other latent Gaussian or latent exponential family models that can be implemented via data augmentation.

A different perspective on stochastic search variable selection approaches is to consider them as a possibility for implementing Bayesian model averaging. Since so far most
applications and implementations of Bayesian model averaging are restricted to linear or generalized linear models, our approach offers a necessary extension of Bayesian model
averaging implementations to a much broader model class. As shown in Section \ref{sec:sim}, it offers improved prediction accuracy and allows for a principled inclusion of
uncertainty about term selection and model structure into inferential statements.


\section*{COMPUTATIONAL DETAILS}
 The approach described and evaluated in this paper is implemented
 in the R-package \blind{\texttt{spikeSlabGAM}} \blind{\citep{spikeSlabGAMJSS}}.

 \section*{ACKNOWLEDGEMENTS}
\blind{ We are indebted
 to Franziska Ru\"{e}ff for letting us
 use the insect allergy data set as an application example and
 to Wolfgang Hartl for letting us use the sepsis survival data.
 Financial support from the German Science Foundation (grant FA
 128/5-1) is gratefully acknowledged. We thank two referees for their constructive comments which helped
 to substantially improve the paper.}

\appendix

\section{MCMC algorithm}

\begin{algorithm}[t]
\begin{algorithmic}[1]
 \STATE Initialize $\bm\tau^{2(0)}, \bm\gamma^{(0)}, {\sigma^2}^{(0)}, w^{(0)}$ and $\bm\beta^{(0)}$ ($\bm\beta^{(0)}$ via
IWLS for non-Gaussian response)
 \STATE Compute $\bm\alpha^{(0)}, \bm\xi^{(0)}, \bm X^{(0)}_{\alpha}$
 \FOR{iterations $t=1,\dots,T$}
    \FOR{blocks $b=1,\dots,b_\alpha$}
        \STATE update $\bm\alpha^{(t)}_b$ from its fcd (Gaussian case, see \eqref{F:fcdCoef})/ via P-IWLS
    \ENDFOR
    \STATE set $\bm X^{(t)}_\xi = \bm X \operatorname{diag}(\operatorname{blockdiag}({\bm
1_d}_1,\dots,{\bm 1_d}_p)\bm\alpha^{(t)})$
    \STATE update $\bm m^{(t)}$ from their fcd: $P(m^{(t)}_l=1|\cdot)=\tfrac{1}{1+\exp(-2\xi^{(t)}_l)},\;l=1,\dots,q$
    \FOR{blocks $b=1,\dots,b_\xi$}
        \STATE update $\bm\xi^{(t)}_b$ from its fcd (Gaussian case, see \eqref{F:fcdCoef})/ via P-IWLS
    \ENDFOR
    \FOR{model terms $j=1,\dots,p$}
        \STATE rescale $\bm\xi^{(t)}_j$ and $\alpha^{(t)}_j$
    \ENDFOR
    \STATE set $\bm X^{(t)}_\alpha = \bm X \operatorname{blockdiag}(\bm\xi^{(t)}_1,\dots,\bm\xi^{(t)}_p)$
    \STATE update ${\tau_1}^{2(t)},...,{\tau_p}^{2(t)}$ from their fcd: $\tau^{2(t)}_j|\cdot \sim \Gamma^{-1}\left(a_\tau+ 1/2,  b_\tau + \tfrac{\alpha_{j}^{2(t)}}{2\gamma^{(t)}_j}\right)$
    \STATE update ${\gamma_1}^{(t)},...,{\gamma_p}^{(t)}$ from their fcd: $\tfrac{P(\gamma^{(t)}_j=1|\cdot)}{P(\gamma^{(t)}_j=v_0|\cdot)} =  v_0^{1/2} \exp\left(\tfrac{(1-v_0)}{2v_0}\tfrac{\alpha_{j}^{2(t)}}{\tau^{2(t)}_j}\right)$
    \STATE update $w^{(t)}$ from its fcd:\\ $w^{(t)}|\cdot \sim \operatorname{Beta}\left(a_w + \sum_j^p \delta_1(\gamma^{(t)}_j),  b_w + \sum_j^p \delta_{v0}(\gamma^{(t)}_j)\right)$
    \IF{$\bm y$ is Gaussian} \STATE update ${\sigma^2}^{(t)}$ from its fcd: ${\sigma^2}^{(t)}|\cdot \sim  \Gamma^{-1}\left(a_{\sigma^2} + n/2,  b_{\sigma^2} + \tfrac{\sum^n_i(y_i-\eta^{(t)}_i)^2}{2}\right) $ \ENDIF
 \ENDFOR
\end{algorithmic}
\caption{MCMC sampler for peNMIG} \label{MCMCalgFCD}
\end{algorithm}

\section{Problems of the Conventional NMIG Prior when Selecting Coefficient Blocks}\label{sec:NMIGProblemApp}

Previous approaches for Bayesian variable selection have primarily concentrated on selection of single coefficients
\citep{George:McCulloch:1993, Ishwaran:2005} or used very low dimensional bases for the representation of smooth effects.
E.g.~\citet{Cottet:Kohn:Nott:2008} use a pseudo-spline representation of their cubic smoothing spline bases with only 3 to 4
basis functions.
In the following, we argue that conventional blockwise Gibbs
sampling is ill suited for updating the state of the Markov chain
when sampling from the posterior of an NMIG model even for
moderately large coefficient blocks. We show that mixing for
$\gamma_j$ will be very slow for blocks of coefficients $\bm
\beta_j$ with $d_j \gg 1$. We suppress the index $j$ in the
following.

\begin{figure}[!ht]
\begin{center}
  \includegraphics[width=\textwidth]{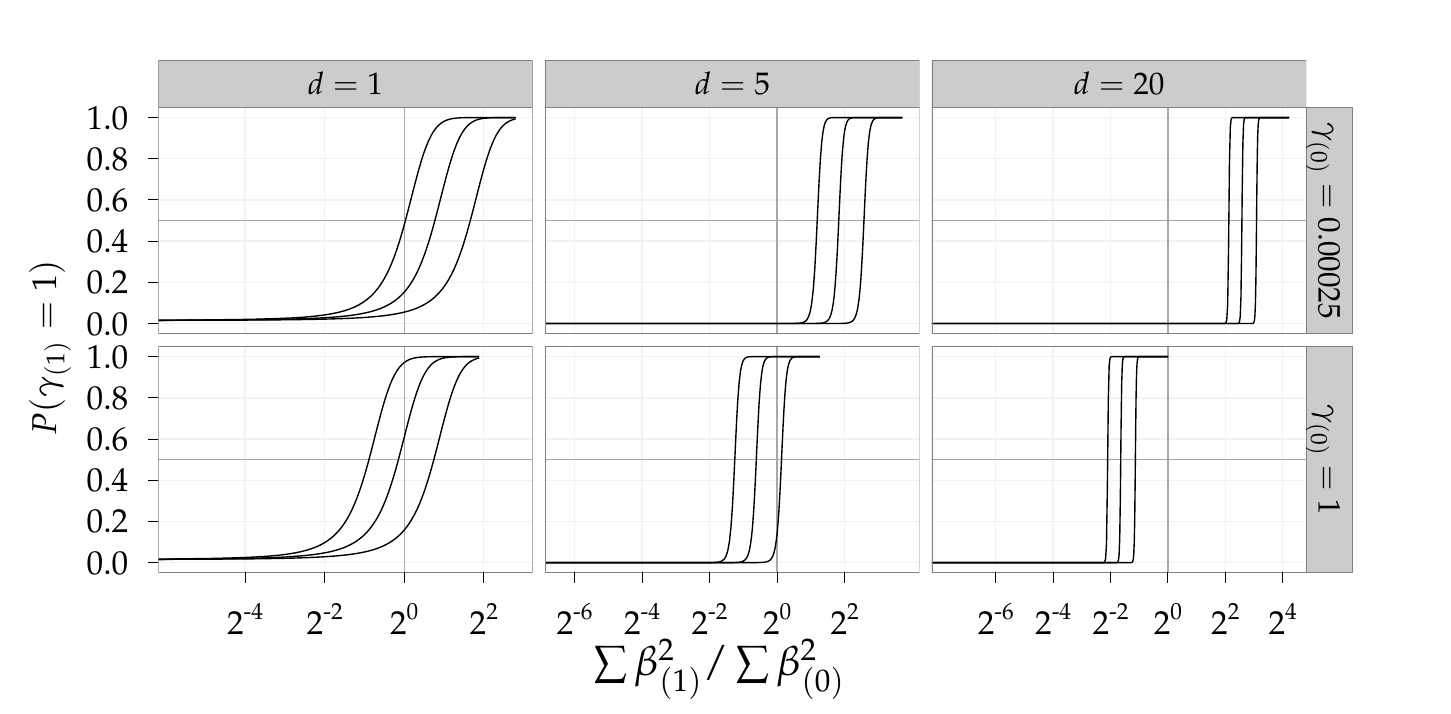}
  \caption[$P(\gamma)$ vs. change in
  $\sum^d\beta^2$]{$P(\gamma)$ as a function of the relative change in
  $\sum^d\beta^2$ for varying $d, \gamma_{(0)}$: Inclusion probability in
 iteration $(1)$ as a function of the ratio between the sum of squared
 coefficients in iteration $(1)$ and $(0)$. Lines in each panel correspond to
 $\tau^2_{(1)}$ equal to the
median of its full conditional and the .1- and .9-quantiles. Upper
row is for $\gamma_{(0)}=1$, lower row for $\gamma_{(0)}=v_0$.
Columns correspond to $d=1,\; 5,\; 20$. Fat gray grid lines denote
inclusion probability $= .5$ and ratio of coefficient sum of squares
$= 1$}
  \label{plotGamma2}
\end{center}
\end{figure}
The following analysis shows that, even if the blockwise sampler is
initially in an ideal state for switching between the spike and the
slab parts of the prior, i.e.~a parameter constellation so that the
full conditional probability $P(\gamma=1|\cdot)=.5$, such a switch
is very unlikely in subsequent iterations for coefficient vectors
with more than a few entries given the NMIG prior hierarchy.

Assume that the sampler starts out in iteration $(0)$ with a
parameter configuration of $a_t, b_t, v_0, w, \tau_{(0)}^2$ and
$\beta_{(0)}$ so that $P(\gamma_{(0)}=1|\cdot)=.5$.  We set $w=.5$.
The parameters for which $P(\gamma=1|\cdot) = .5$  satisfy the
following relations:
\begin{align*}
\frac{P(\gamma=1|\cdot)}{P(\gamma=v_0|\cdot)} &=  v_0^{d/2}
 \exp\left(\frac{(1-v_0)}{2v_0}\frac{\sum^d \beta^2}{\tau^2}\right)  = 1, \\
\intertext{so that $P(\gamma=1|\cdot)>.5$ if} \frac{\sum^d
\beta^2}{d\tau^2} &> -\frac{v_0}{1-v_0}\log(v_0).
\end{align*}
Assuming a given value $\tau^2_{(0)}$, set
\begin{align*}
\sum^d\beta^2_{(0)}= -\frac{d v_0}{1-v_0} \log(v_0)\tau^2_{(0)}.
\end{align*}
Now $\gamma_{(0)}$ takes on both values $v_0$ and 1 with equal
probability, conditional on all other parameters.

In the following iteration, $\tau^2_{(1)}$ is drawn from its full
conditional $\Gamma^{-1}(a_t + d/2, b_t +
\frac{\sum^d\beta^2_{(0)}}{2\gamma_{(0)}})$. Figure \ref{plotGamma2}
shows $P(\gamma_{(1)} = 1| \tau^2_{(1)}, \sum^d\beta^2_{(1)})$ as a
function of $\sum^d\beta^2_{(1)}/\sum^d\beta^2_{(0)}$ for various
values of $d$. The 3 lines in each panel correspond to
$P(\gamma_{(1)} = 1| \tau^2_{(1)}, \sum^d\beta^2_{(1)})$ for values
of $\tau^2_{(1)}$ equal to the median of its full conditional as
well as the .1- and .9-quantiles. The lower row in the Figure plots
the function for $\gamma_{(0)}=1$, the upper row for
$\gamma_{(0)}=v_0$.

So, if we start in this ``equilibrium state'' we begin iteration
$(0)$ with $v_0, w$, $\tau_{(0)}^2$, and $\bm \beta_{(0)}$ so that
$P(\gamma_{(0)}=1|\cdot)=0.5$. We then determine
\mbox{$P(\gamma_{(1)} \neq \gamma_{(0)}| \tau^2_{(1)},
\sum^d\beta^2_{(1)})$} as a function of
$\sum^d\beta^2_{(1)}/\sum^d\beta^2_{(0)}$ for
\begin{itemize}
\item various values of $\operatorname{dim}(\bm\beta_j)=d$,
\item $\gamma_{(0)}=1$ and $\gamma_{(0)}=v_0$,
\item $\tau^{2}_{(1)}$ at the $.1$, $.5$, $.9$-quantiles of its
 conditional distribution given $\bm \beta_{(0)}, \gamma_{(0)}$.
\end{itemize}

The leftmost column in Figure \ref{plotGamma2} shows that moving
between $\gamma=1$ and $\gamma = v_0$ is easy for $d=1$: For a large
range of realistic values for
$\sum^d\beta^2_{(1)}/\sum^d\beta^2_{(0)}$, moving back to
$\gamma_{(1)}= v_0$ from $\gamma_{(0)} = 1$ (lower panel) has
reasonably large probability, just as moving from $\gamma_{(0)} =
v_0$ to $\gamma_{(1)}= 1$ (lower panel) is fairly likely for
realistic values of $\sum^d\beta^2_{(1)}/\sum^d\beta^2_{(0)}$. For
$d=5$, however, $P(\gamma_{(1)} = 1| \cdot)$ already resembles a
step function. For $d=20$, if
$\sum^d\beta^2_{(1)}/\sum^d\beta^2_{(0)}$ is not smaller than
$0.48$, the probability of moving from $\gamma_{(0)} = 1$ to
$\gamma_{(1)} = v_0$ (lower panel) is practically zero for 90\% of
the values drawn from $p(\tau^2_{(1)}|\cdot)$. However, draws of
$\beta$ that reduce $\sum^d\beta^2$ by more than a factor of $0.48$
while $\gamma=1$ are unlikely to occur in real data.
It is also extremely unlikely to move back to $\gamma_{(1)} = 1$
when $\gamma_{(0)} = v_0$, unless
$\sum^d\beta^2_{(1)}/\sum^d\beta^2_{(0)}$ is larger than $2.9$.
Since the full conditional for $\beta$ is very concentrated if
$\gamma = v_0$, such moves are highly improbable and correspondingly
the sampler is unlikely to move away from $\gamma = v_0$. Numerical
values for the graphs in Figure \ref{plotGamma2} were computed for
$a_\tau=5,\; b_\tau=50,\; v_0=0.00025$ but similar problems arise
for all suitable hyperparameter configurations.

In summary, mixing of the indicator variables $\gamma$ will be very
slow for long subvectors. In experiments, we observed posterior
means of $P(\gamma=1)$ to be either $\approx  0$ or $\approx 1$
across a wide variety of settings, even for very long chains,
largely depending on the starting values of the chains. A
multiplicative parameter expansion offers a possible remedy, with
the added benefit of inducing very desirable shrinkage properties
for the resulting estimates as shown in the article.

\section{Simulation Results}\label{sec:simApp}
In the following Sections \ref{sec:simGAM:Gauss} and
\ref{sec:simGAM:Pois}, we compare the performance of \mbox{peNMIG}
in (generalized) additive models (GAMs) as implemented in package \package{spikeSlabGAM} \blind{\citep{spikeSlabGAMJSS}} to that of  component-wise
boosting \citep{mboost} in terms of predictive MSE and complexity
recovery. As a reference, we also fit a conventional GAM (as
implemented in \package{mgcv} \citep{Wood:2008}) based on the
``true'' formula (i.e. a model without any of the ``noise'' terms),
which we subsequently call the ``oracle''-model. For Gaussian
responses only, we also compare our results to those from ACOSSO
\citep{Storlie:Bondell:Reich:2010}. ACOSSO is not able to fit
non-Gaussian responses. Section \ref{sec:simGAM:conc} investigates
the effects of concurvity on effect estimates and term selection
for our method and those of some recently proposed competitors.

We supply separate base learners for the linear and smooth parts of
covariate influence for the component-wise boosting in order to
compare complexity recovery between boosting and our approach. We
use 10-fold cross validation on the training data to determine the
optimal stopping iteration for \package{mboost} and count a
baselearner as included in the model if it is selected in at least
half of the cross-validation runs up to the stopping iteration. BIC
is used to determine the tuning parameter for ACOSSO. We did not compare our
approach to \citet{Reich:Storlie:Bondell:2009}, which is implemented for Gaussian responses,
since the available \textsf{R} implementation is impractically slow --
computation times are usually 15 - 30 times those of our peNMIG implementation.

For both Gaussian responses (Section \ref{sec:simGAM:Gauss}) and Poisson
responses (Section \ref{sec:simGAM:Pois}), the data generating process
has the following structure:
\begin{itemize}
  \item We define 4 functions
   \begin{itemize}
    \item $f_1(x) = x$,
    \item $f_2(x) = x + \frac{(2x-2)^2}{5.5}$,
    \item $f_3(x)= -x + \pi\sin(\pi x)$,
    \item $f_4(x)= 0.5x + 15\phi(2(x-.2)) -
        \phi(x+0.4)$, where $\phi()$ is the standard normal density function,
   \end{itemize}
   which enter into the linear predictor. Note that all of them have (at least)
   a linear component.
  \item We define 2 scenarios:
  \begin{itemize}
    \item a ``low sparsity'' scenario: Generate 16 covariates, 12 of which have
        non-zero influence. The true linear predictor is $\eta = f_1(x_1) +
        f_2(x_2) + f_3(x_3) + f_4(x_4) + 1.5(f_1(x_5) + f_2(x_6) + f_3(x_7) +
        f_4(x_8)) + 2(f_1(x_9) + f_2(x_{10}) + f_3(x_{11}) +
        f_4(x_{12}))$.
    \item a ``high sparsity'' scenario: Generate 20 covariates, only 4 of which have
        non-zero influence and  $\eta = f_1(x_1) + f_2(x_2) + f_3(x_3) +
        f_4(x_4)$.
  \end{itemize}
  \item The covariates are either
   \begin{itemize}
    \item  $\stackrel{\iid}{\sim}U[-2,2]$ or
    \item  from an AR(1) process with correlation $\rho=0.7$.
   \end{itemize}
   \item We simulate 50 replications for each combination of the various
   settings.
\end{itemize}
We compare 9 different prior specifications arising from the combination of
\begin{itemize}
  \item $(a_\tau, b_\tau) = (5, 25)$, $(10, 30)$, $(5, 50)$
  \item $v_0= 0.00025, 0.005, 0.01$
\end{itemize}
Predictive MSE is evaluated on test data sets with 5000
observations. Complexity recovery, i.e. how well the different
approaches select covariates with true influence on the response and
remove covariates without true influence on the response is measured
in terms of accuracy, defined as the number of correctly classified
model terms (true positives and true negatives) divided by the total
number of terms in the model. For example, the full model in the
``low sparsity'' scenario has 32 potential terms under selection
(linear terms and basis expansions/smooth terms for each of the 16
covariates), only 21 of which are truly non-zero (the linear terms
for the first 12 covariates plus the 9 basis expansions of the
covariates not associated with the linear function $f_1()$).
Accuracy in this scenario would then be determined as the sum of the
correctly included model terms plus the correctly excluded model
terms, divided by 32.

\subsection{Gaussian response}\label{sec:simGAM:Gauss}
\begin{figure}[!tbp]
\begin{center}
  \includegraphics[width=\textwidth]{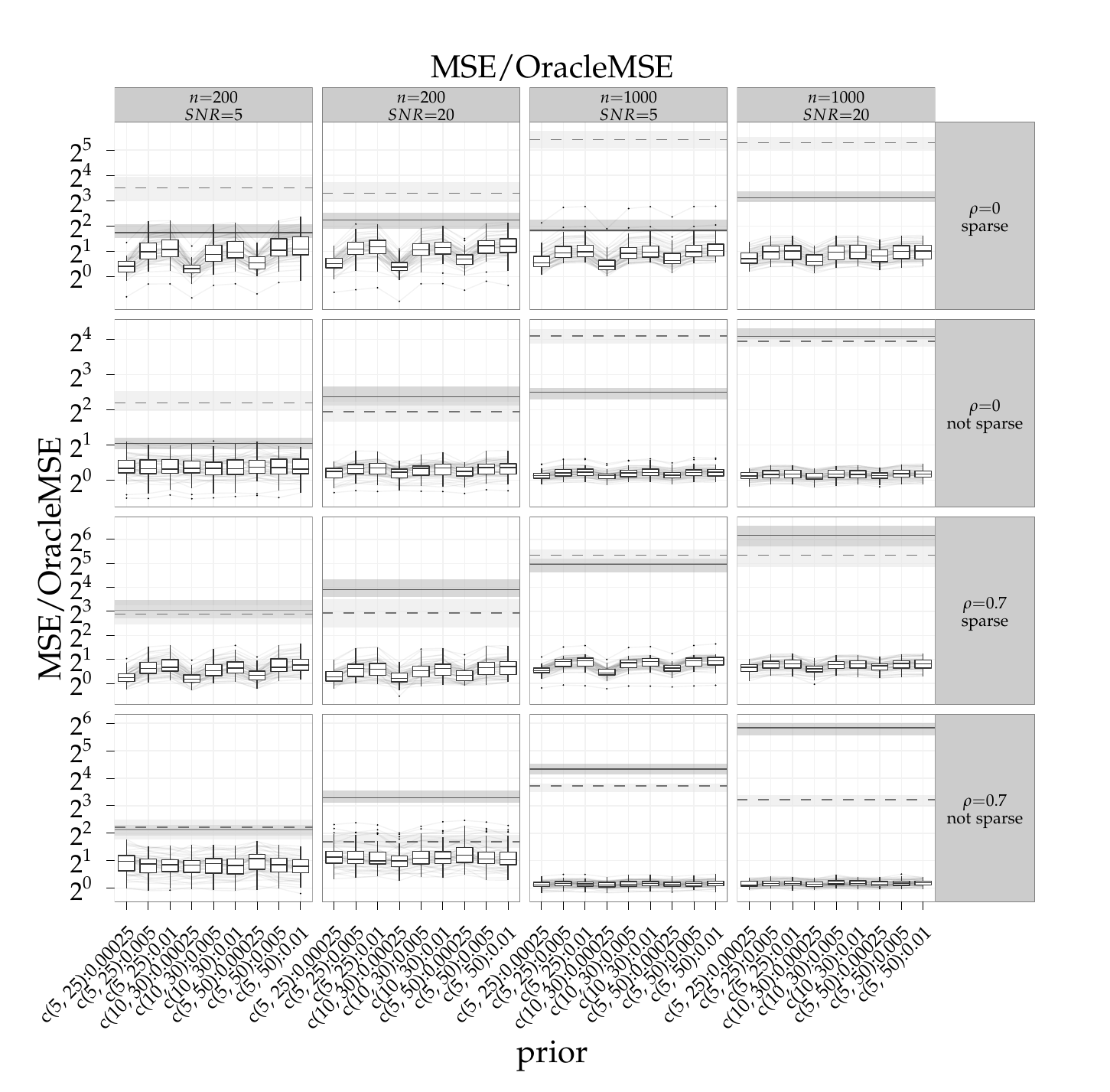}
  \caption[Gaussian AM: Relative predictive MSE]{Prediction MSE divided by
  oracle MSE for Gaussian response. Boxplots show results for the different prior settings,
   horizontal ribbons show results for \package{mboost} (solid) and ACOSSO
   (dashed), respectively: Shaded region gives IQR, line represents median. Dark
   grey lines connect results for the same replication.
   Columns from left to right: 200 obs. with SNR=5, 20; 1000 obs. with SNR=5, 20. Rows from top to bottom: uncorrelated obs. with sparse and unsparse predictor, correlated obs. with sparse and unsparse predictor. Vertical axis
   is on binary log scale.}
  \label{mdamGauss_relMSE}
\end{center}
\end{figure}
\begin{figure}[!tbp]
\begin{center}
  \includegraphics[width=\textwidth]{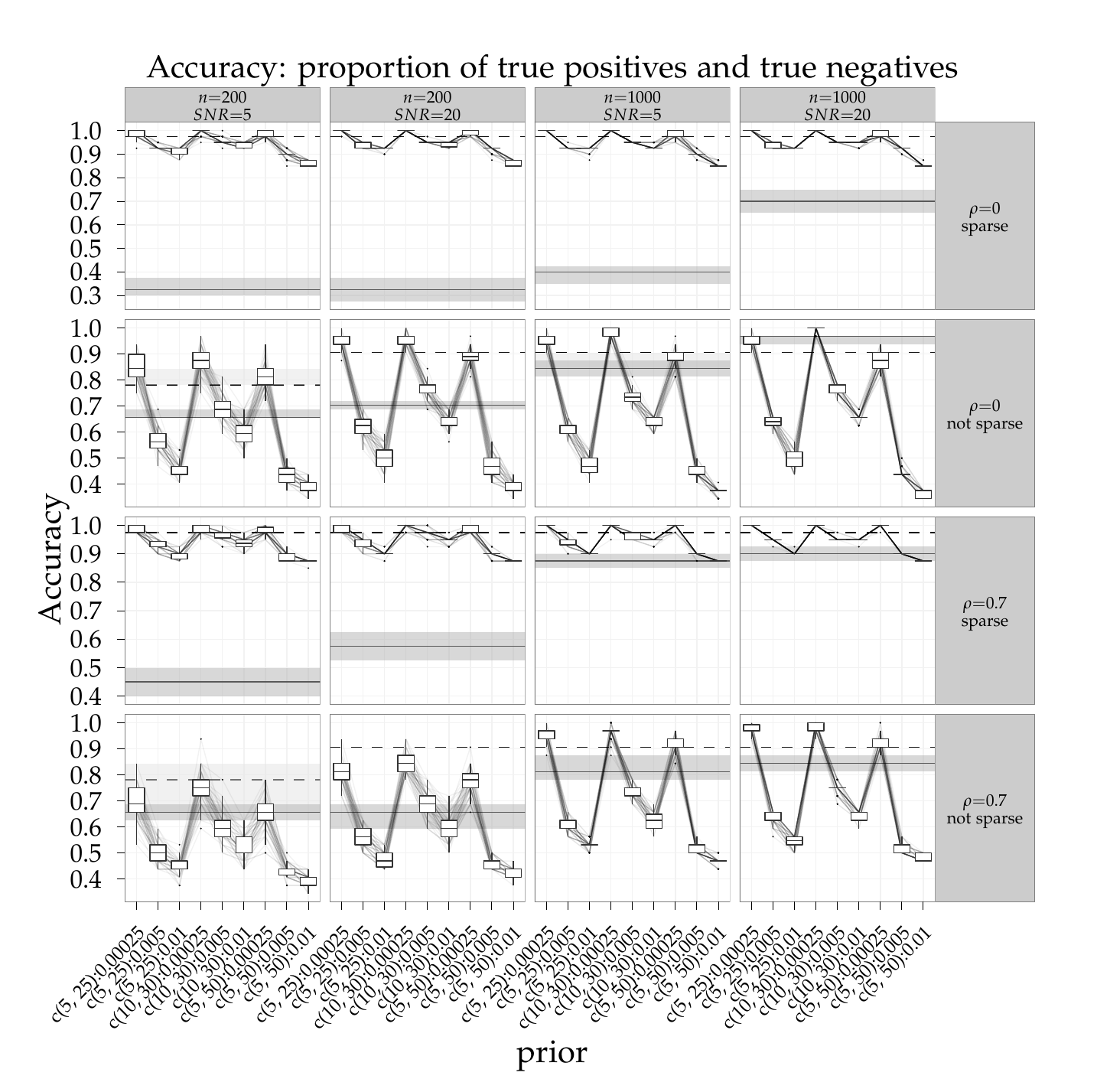}
  \caption[Gaussian AM: Complexity recovery]{Complexity recovery for
  Gaussian response: proportion of correctly included and excluded model terms.
   Boxplots show results for the
  different prior settings, horizontal ribbons show results for
  \package{mboost} (solid) and ACOSSO (dashed), respectively: Shaded region
  gives IQR, line represents median. Dark grey lines connect results for the same replication. Columns from left to right: 200
   obs. with SNR=5, 20; 1000 obs. with SNR=5, 20. Rows from top to bottom: uncorrelated obs. with sparse and unsparse
   predictor, correlated obs. with sparse and unsparse predictor.}
  \label{mdamGauss_sens}
\end{center}
\end{figure}
In addition to the basic structure of the data generating process
described at the beginning of this section, the data generating
process for the Gaussian responses has the following properties:
\begin{itemize}
    \item signal-to-noise-ratio $\operatorname{SNR} = 5, 20$
    \item number of observations: $n = 200, 1000$
\end{itemize}

Figure \ref{mdamGauss_relMSE} shows the mean squared prediction error divided by the one achieved by the ``oracle''-model, a
conventional GAM without any of the noise variables. Predictive performance is very robust against the different prior settings
especially for the settings with low sparsity. Different prior settings also behave similarly within replications, as shown by
the mostly parallel grey lines. Predictions are more precise than those of both boosting and ACOSSO, and this improvement in
performance relative to the ``true'' model is especially marked for $n=1000$ (two rightmost columns). With the exception of the
first scenario, the median relative prediction MSE is $< 2$ everywhere, while both boosting and ACOSSO have a median relative
prediction MSE above 4 in most scenarios that goes up to above 32 and 64 for ACOSSO and boosting, respectively, in the ``large
sample, correlated covariates'' cases. In the ``large sample, low sparsity'' scenarios (two leftmost columns in rows two and
four), the performance of our approach comes very close that of the oracle model -- the relative prediction MSEs are close to
one.

Figure \ref{mdamGauss_sens} shows the proportion of correctly included and excluded terms (linear terms and basis expansions)
in the estimated model. Except for $v_0=0.00025$, accuracy is consistently lower than for ACOSSO. However, a direct comparison
with ACOSSO is not entirely appropriate because ACOSSO does not differentiate between smooth and linear terms, while
\package{mboost} and our approach do. Therefore ACOSSO solves a less difficult problem. Estimated inclusion probabilities are
very sensitive to $v_0$ and comparatively robust against $(a_\tau, b_\tau)$. Across all settings, $v_0=0.00025$ delivers the
most precise complexity recovery, with sensitivities consistently above $0.7$.  The accuracy of peNMIG is better than
\package{mboost} for the sparse settings (first and third rows) because the specificity of our approach is $>.97$ across
settings, regardless of the prior (!), while \package{mboost} mostly achieves only very low specificity, but fairly high
sensitivity.

\subsection{Poisson response}\label{sec:simGAM:Pois}
\begin{figure}[!tbp]
\begin{center}
  \includegraphics[width=\textwidth]{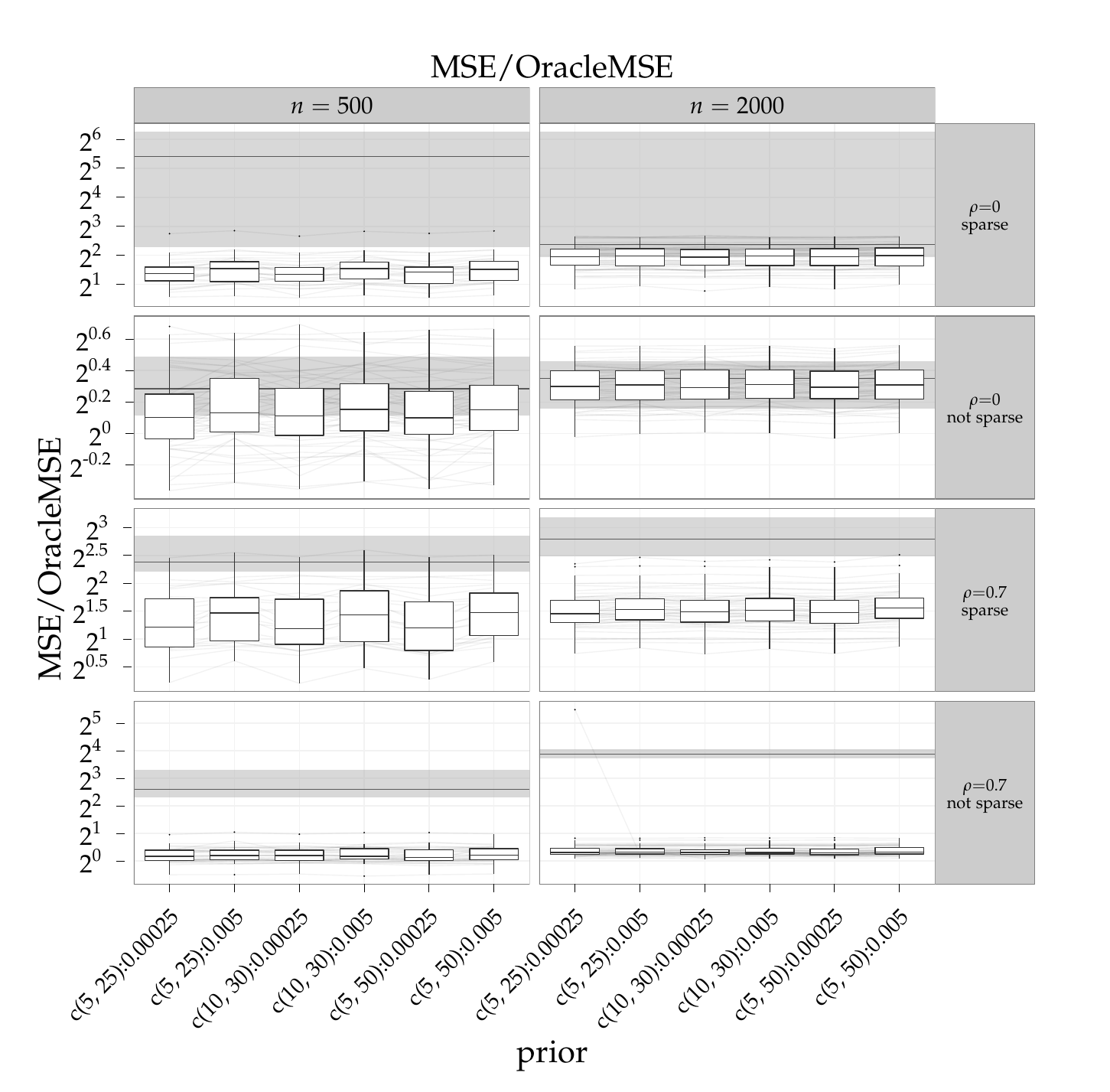}
  \caption[Poisson GAM: Relative predictive MSE]{Prediction MSE divided by
  oracle MSE (on the scale of the linear predictor). Boxplots show results for the different prior
  settings. Horizontal ribbons show results for \package{mboost}:
   shaded region gives IQR, line represents median. Dark grey lines
   connect results for the same replication.  Columns from left to
   right: 500 obs., 2000 obs. Rows from top to bottom: uncorrelated obs. with sparse and unsparse
   predictor, correlated obs. with sparse and unsparse predictor. Vertical axis
   is on binary log scale.}
  \label{mdamPoisson_relMSE}
\end{center}
\end{figure}
\begin{figure}[!tbp]
\begin{center}
  \includegraphics[width=\textwidth]{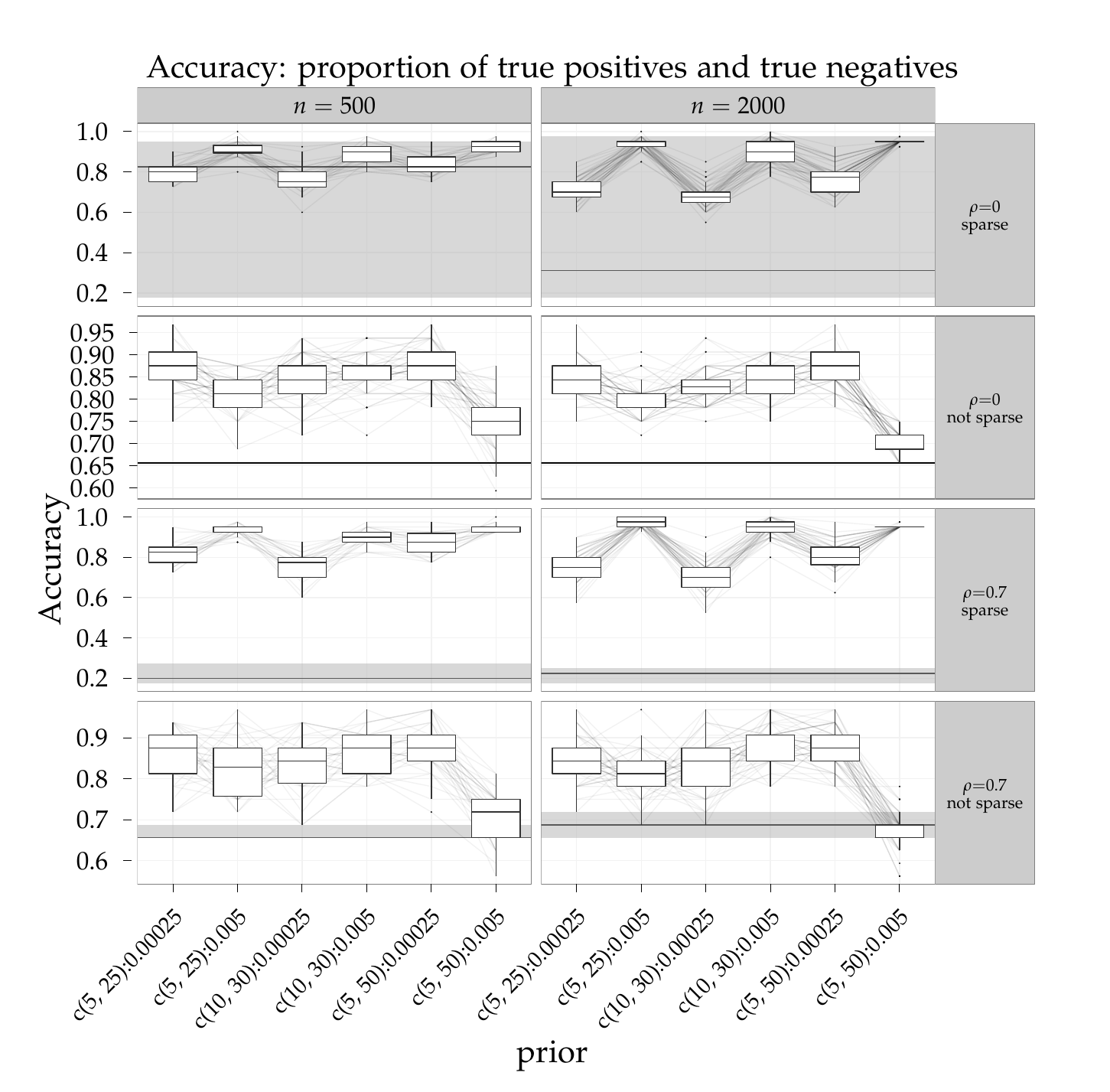}
  \caption[Poisson GAM: Complexity recovery]{Complexity recovery for
  poisson response: proportion of correctly included and excluded model terms. Boxplots show results for the
  different prior settings. Horizontal ribbons show results for
  \package{mboost}: shaded region gives IQR, line represents median. Dark grey lines connect
   results for the same replication.
   Columns from left to right: 500 obs., 2000 obs.
   Rows from top to bottom: uncorrelated obs. with sparse and unsparse
   predictor, correlated obs. with sparse and unsparse predictor.}
  \label{mdamPoisson_acc}
\end{center}
\end{figure}
In addition to the basic structure of the data generating process
described at the beginning of this section, the data generating
process for the Poisson responses has the following properties:
\begin{itemize}
    \item number of observations: $n = 500, 2000$
    \item responses are generated with overdispersion:\\
    \mbox{$y_i \sim Po\left(s_i \exp(\eta_i)\right);\; s_i \sim U[0.66, 1.5]$}
\end{itemize}
We did not use $v_0= 0.01$ for this experiment because of its
inferior performance in terms of complexity recovery in the Gaussian
case.

Figure \ref{mdamPoisson_relMSE} shows the mean squared prediction
error (on the scale of the linear predictor) divided by the one
achieved by the ``oracle''-GAM that includes only the relevant
covariates and no noise terms. Predictive performance is very robust
against the different prior settings. Different prior settings also
behave similarly within replications, as shown by the mostly
parallel grey lines. Predictions are more precise than those
of\package{mboost}, especially for smaller data sets (left column)
and correlated responses (two bottom rows). For the ``low sparsity,
correlated covariates'' setting (bottom row), the performance of our
approach comes fairly close to that of the ``oracle''-GAM, with
relative prediction errors mostly between 1 and 1.5, and
occasionally even improving on the oracle model for $n=500$.

Figure \ref{mdamPoisson_acc} shows the proportion of correctly
included and excluded terms (linear terms and basis expansions) in
the estimated models. Estimated inclusion probabilities are
sensitive to $v_0$ and comparatively robust against $(a_\tau,
b_\tau)$.  The smaller value for $v_0$ tends to perform better in
the unsparse settings (second and fourth rows) since it forces more
terms into the model (resulting in higher sensitivity and lower
specificity) and vice versa for the sparse setting and the larger
$v_0$. Complexity recovery is (much) better across the different
settings and priors for our approach than for boosting. The constant
accuracy for \package{mboost} in the low sparsity scenario with
uncorrelated responses (second row) is due to its very low
specificity: It includes practically all model terms all the time.

\subsection{Gaussian GAM with concurvity}\label{sec:simGAM:conc}
We use a similar data generating process as the one used in the previous Subsections:
\begin{itemize}
  \item We define functions $f_1(x)$ to $f_4(x)$ as in the data generating process for the previous Subsections.
  \item We use 10 covariates: The first 4 are associated with functions $f_1$ to $f_4$, respectively, while $x_5$ to
  $x_{10}$ are ``noise'' variables without contribution to the linear predictor.
  \item covariates $x$ are $\sim U[-2, 2]$
  \item we distinguish 3 scenarios of concurvity:
  \begin{itemize}
    \item in scenario 1, $x_4 = c\cdot g(x_3) + (1-c)\cdot u$, i.e., two covariates with real influence on the predictor are functionally related.
    \item in scenario 2, $x_5 = c\cdot g(x_4) + (1-c)\cdot u$, i.e., a ``noise'' variable is a noisy version of a function of a covariate with direct influence.
    \item in scenario 3, $x_4 = c\cdot g(x_5) + (1-c)\cdot u$, i.e., a covariate with direct influence is a noisy version of a ``noise'' variable.
  \end{itemize}
  where $g(x)= 2\Phi(x, \mu=-1, \sigma^2=0.16) + 2\Phi(x, \mu=1, \sigma^2=0.09) - 4\phi(x) -2$ with $\Phi(x,\mu,\sigma^2)$ defined as the
  cdf of the respective Gaussian distribution, $\iid$ standard normal variates $u$, and the parameter $c$ controlling the amount of concurvity:
  $c=1$ for perfectly deterministic relationship, and $c=0$ for independence. In our simulation, $c = 0, .2, .4, .6, .8, 1$.\\
  \item we use signal-to-noise ratio SNR$=1, 5$.
  \item we simulate 50 replications for each combination of the various settings.
\end{itemize}
Predictive MSE is evaluated on test data sets with 5000 observations.
We use the default prior $a_\tau=5, b_\tau=25, v_0=0.00025$ for our approach and present results for $12$ chains run in parallel for 2600 iterations
each, discarding the first 100 as burn-in and keeping every fifth iteration.
We compare predictive MSE and the sensitivity of the
selection to various degrees of concurvity to results from
\begin{itemize}
  \item \package{mboost}, as above,
  \item \texttt{BART}: Bayesian additive regression trees \citep{Chipman:George:McCulloch:2010} as implemented in R-package \package{BayesTree} \citep{Chipman:McCulloch:2010},
  as an example of a non-parametric method that does not yield interpretable models,
  \item \texttt{hgam}: the high-dimensional additive model approach of \citet{Meier:Geer:Buehlmann:2009}, as implemented in R-package \package{hgam}
  \item \texttt{spam}: the approach for sparse additive models by \citet{Ravikumar:2009}, with a covariate assumed to be selected as a linear influence if its
    estimated associated degrees of freedom were between 0.1 and 1, and assumed to be selected as a non-linear influence if its
    estimated associated degrees of freedom were $>1$.
  \item \texttt{mgcv-DS}: the double shrinkage approach for GAM estimation and term selection described in \citet{Marra:Wood:2011},
  as implemented in R-package \package{mgcv}, with a covariate assumed to be selected as a linear influence if its
    estimated associated degrees of freedom were between 0.1 and 1, and assumed to be selected as a non-linear influence if its
    estimated associated degrees of freedom were $>1$.
\end{itemize}
\begin{figure}[!tbp]
\begin{center}
  \includegraphics[width=\textwidth]{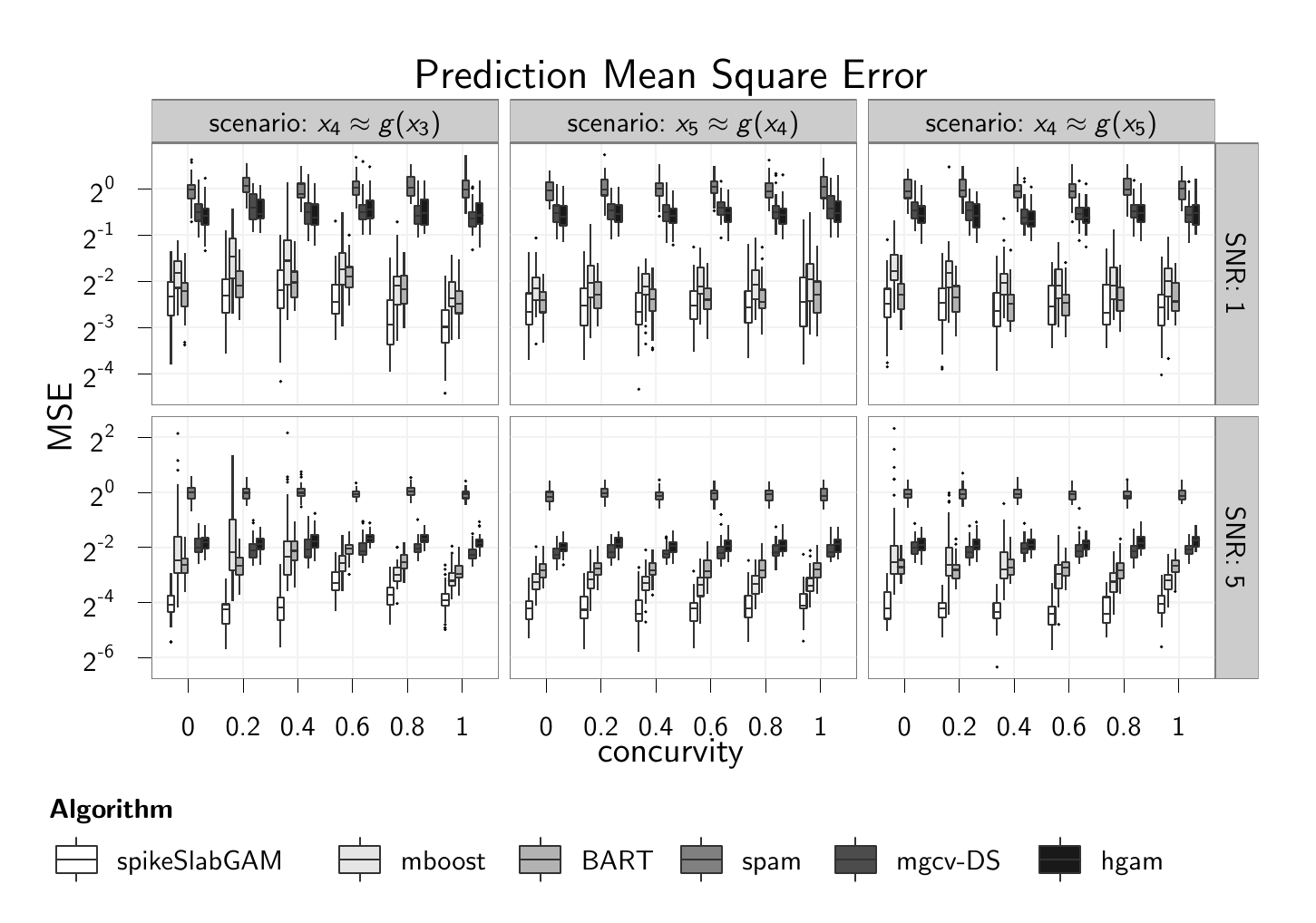}
  \caption[Concurvity AM: predictive MSE]{Prediction MSE. Boxplots show results for the different algorithms.
  Columns for the three scenarios, top row for signal to noise ratio 1, bottom row for signal to noise ratio 5. Vertical axis
   is on binary log scale.}
  \label{mdamConc_MSE}
\end{center}
\end{figure}
Figure \ref{mdamConc_MSE} compares prediction MSEs for the various methods across scenarios and signal-to-noise ratios
for varying degrees of concurvity. It is clear to see that our approach dominates in
terms of prediction accuracy in these difficult settings, with BART and \package{mboost} as fairly close competitors.
\begin{figure}[!tbp]
\begin{center}
 \includegraphics[width=\textwidth]{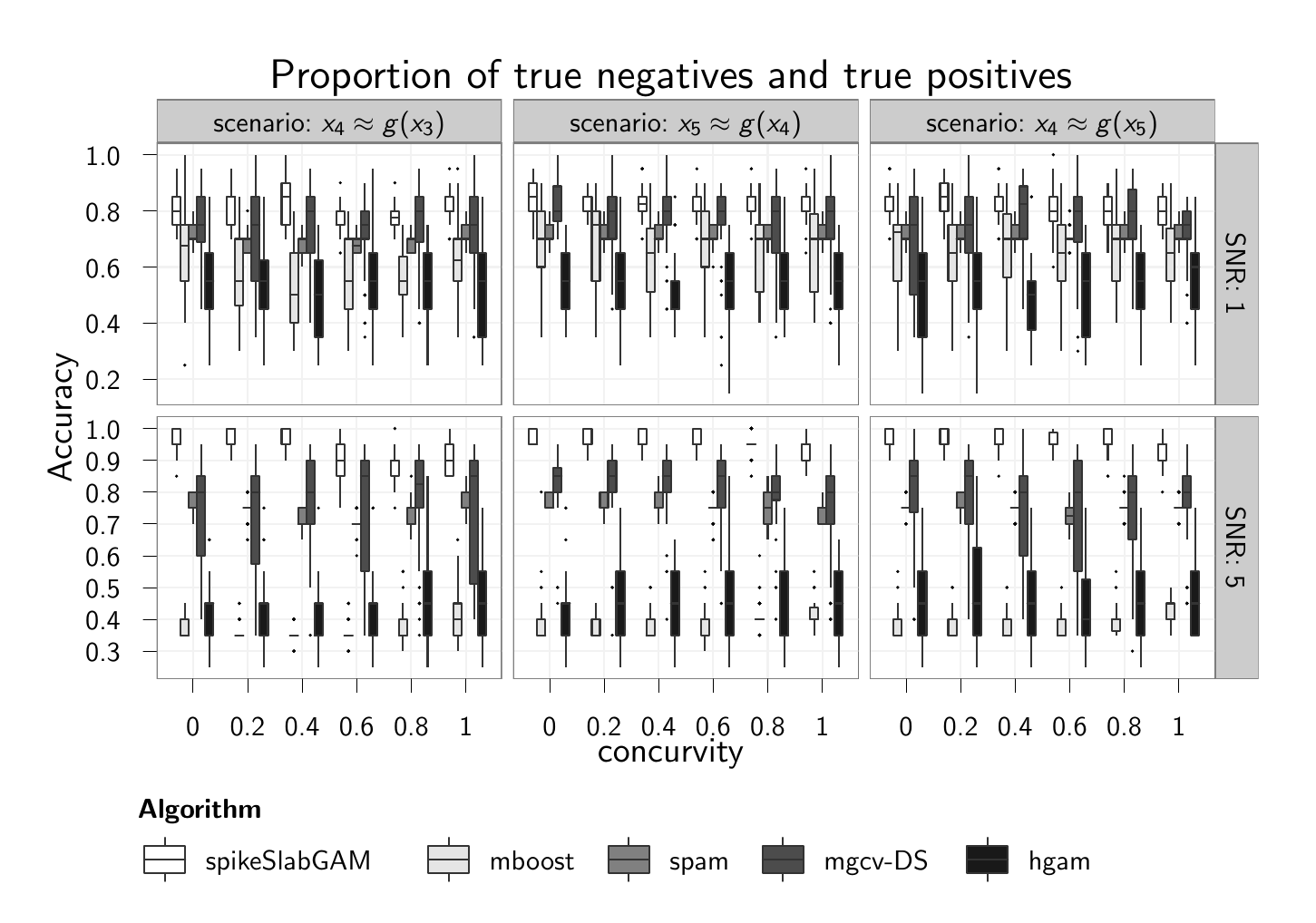}
  \caption[Concurvity AM: Selection Accuracy]{Selection accuracy as proportion of correctly selected or removed predictors.
  Boxplots show results for the different algorithms. Columns for the three scenarios, top row for signal to noise ratio 1, bottom row for signal to noise ratio 5.}
  \label{mdamConc_Acc}
\end{center}
\end{figure}
Figure \ref{mdamConc_Acc} shows the proportion of correctly selected or removed covariates
(i.e.,~``true positives'' and ``true negatives''). As for prediction accuracy, our approach outperforms the rest, with
the double shrinkage approach of \citet{Marra:Wood:2011} a close second for selection accuracy for the noisy setting (but much worse in terms of prediction).
Figure \ref{mdamConc_Acc} does not include BART since its implementation in \package{BayesTree} does not offer clear inclusion or exclusion
indicators. The only variable importance measure (i.e., how often a given variable was used in nodes across the ensemble of trees)
returned by BART had roughly the same mean for influential and noise variables in all our simulations and would not have yielded a
useable picture of the true predictor structure.
\begin{figure}[!tbp]
\begin{center}
 \includegraphics[width=\textwidth]{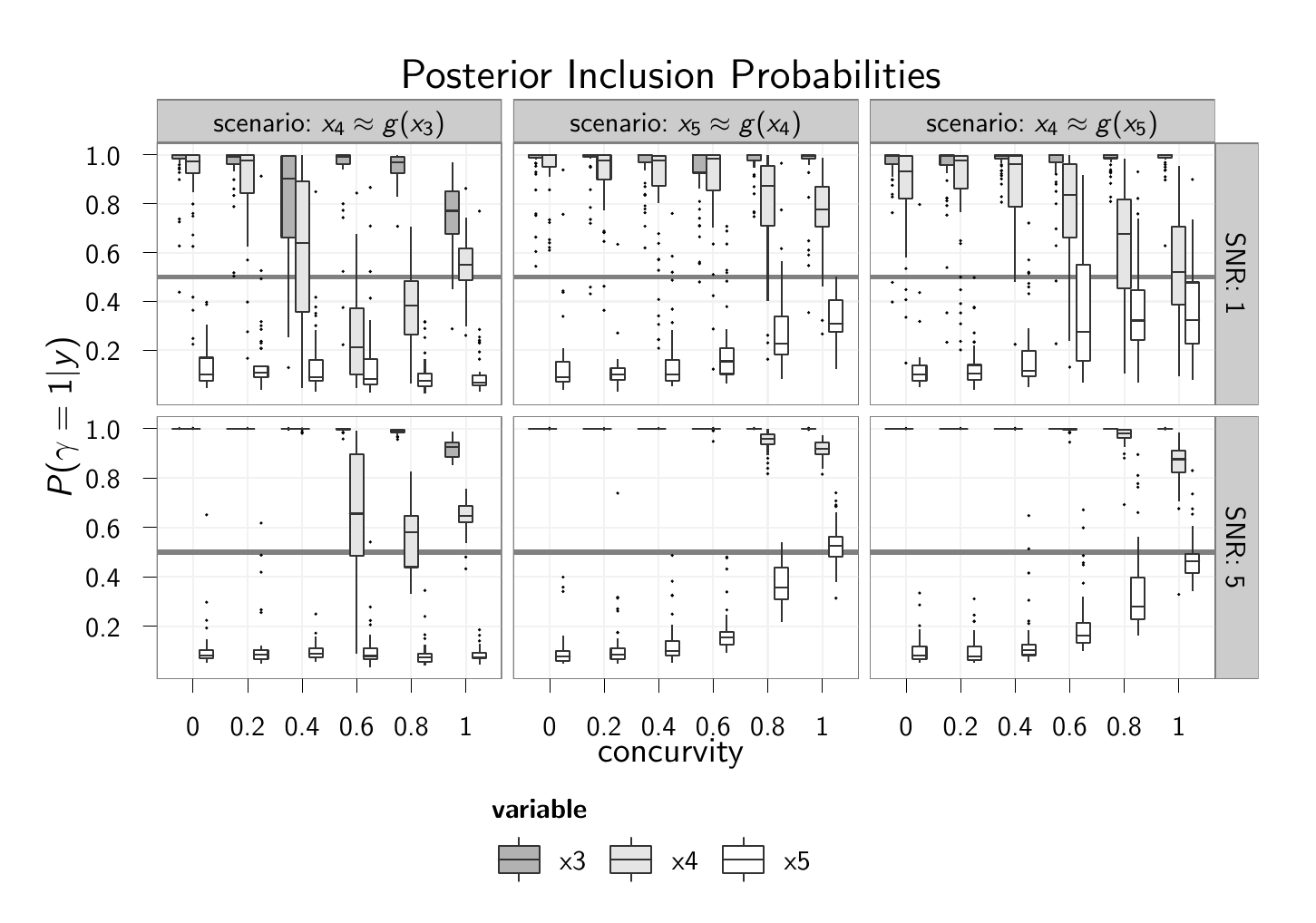}
  \caption[Concurvity AM: $P(\gamma=1|\bm y)$]{Posterior inclusion probabilities for $x_3, x_4, x_5$. Shown are the respective maxima of the posterior inclusion probabilities for
  the linear and smooth effects of each variable in each replication.
  Columns for the three scenarios, top row for SNR 1, bottom row for SNR 5. Fat grey horizontal line denotes $P(\gamma=1)=0.5$}
  \label{mdamConc_pInc}
\end{center}
\end{figure}
Finally, figure \ref{mdamConc_pInc} shows that estimated inclusion probabilities are fairly unreliable
for intermediate to strong concurvity in noisy settings
(top row) if both variables that are involved have separate effects (left column) -- in this scenario, estimated inclusion probabilities
 of the covariate $x_3$, which is a (noisy) function of another covariate $x_4$, decrease dramatically for intermediate concurvity
and recover somewhat for perfect curvilinearity, where inclusion probabilities for $x_3$ are somewhat reduced.
Note however, that if a model including interactions of $x_3$ and $x_4$ is specified in this scenario, our approach will often
select those. This behavior makes sense if the effects of $x_3$ and $x_4$ cannot be clearly separated, as in this case.\\
If a variable $x_5$ without true effect is a (noisy) function of another covariate $x_4$ with true effect (second column),
inclusion probabilities are fairly stable
and yield correct inferences about the model structure unless the curvilinear relationship is perfect (concurvity$=1$),
at which point it again becomes impossible to disentangle the effect of $x_4$ and $x_5$. Even so, using a
threshold for inclusion of $P(\gamma=1|\bm y) > 0.5$  the correct model structure would have been identified in 44 out of 50 replicates for
low SNR and 17 out of 50 for high SNR.
For noisy data (top row), the scenario in which a noisy version $x_4$ of a spurious covariate $x_5$ has true effects (third column)
is problematic,  with inclusion probabilities for $x_4$ decreasing strongly under intermediate and strong concurvity.
Note, however, that even for strong concurvity $\geq 0.6$, the true model structure was identified at least 24 times out
of 50 replicates for low SNR and at least 41 times out of 50 replicates for high SNR.

\subsection{Summary}
The simulations for generalized additive models show that the
proposed peNMIG-Model is very competitive in terms of estimation
accuracy and confirms that estimation results are robust against
different hyperparameter configurations even in fairly complex
models, and under strong concurvity. Model selection is more sensitive towards hyperparameter
configurations, especially $v_0$. For smaller $v_0$,
\package{spikeSlabGAM} seems to be able to distinguish between important
and irrelevant terms fairly reliably.

The performance of peNMIG as implemented in
\package{spikeSlabGAM} seems to be very competitive to that of
component-wise boosting as implemented in \package{mboost} and clearly dominates
other function selection approaches in our concurvity simulation study.
Simulation results for an earlier, more rudimentary implementation
of the peNMIG model on identical data generating processes and for
many other settings are published in \blind{\citet{Scheipl:2010}}.

\section{Additional Case Studies}\label{sec:CaseStudiesApp}
According to German law, increases in rents for flats can be justified based on ``average rents'' paid for flats that are
comparable in size, equipment, quality and location. As a consequence, most larger cities publish rental guides that provide
such average rents, obtained from regression models with net rents or net rents per square meter as dependent variables.

\paragraph{Data:}
We analyze data on approximately 3,000 flats in Munich collected by Infratest Sozialforschung for the 2007 rental guide. The
original data contain approximately 270 covariates describing characteristics of the flats as diverse as the quality of
bathroom equipment, whether the flat is rented for the first time, the presence and size of a garden or a balcony, etc. While
most of the covariates are categorical, there are some continuous covariates of designated interest that are suspected to have
nonlinear impact on the net rent per square meter. Moreover, a spatial variable is available that represents the subdistrict of
Munich where the flat is located.

\paragraph{Model:} We will model net
rents per square meter with a high-dimensional geoadditive
regression with Gaussian errors and linear predictor
\begin{align*}
\bm \eta_i = \beta_0 + f_{\text{spat}}(s_i) + \sum^{267}_{j=1}
f(x_{ij}),
\end{align*}
where $f_{\text{spat}}(s_i)$ is the spatial effect of subdistrict $s_i$ modeled with a Gaussian Markov random field (GMRF). The
linear predictor additionally contains potential smooth effects of the year of construction, floorspace and begin of tenancy as
well as 265 other potentially influential and mostly categorical covariates $\bm x_j$. In total, this model has 594
coefficients in 269 model terms. Term selection is particularly challenging in this scenario because the available covariates
include a number of redundant and highly collinear covariates such as an indicator variable for the presence of at least
one balcony, a numeric variable giving the size of the flat's balconies in square meters and a count variable giving the number
of balconies.

Hyperparameters were set to the default values determined in the simulation studies, i.e. $a_\tau=5, b_\tau=25, v_0=0.00025$
and $a_w=b_w=1$. Estimates are based on 20 parallel chains running for 2500 iterations each after a burn-in of 500 iterations,
with every fifth iteration saved.



\paragraph{Results:}

\begin{figure}[!htbp] \centering
\includegraphics[width=\textwidth]{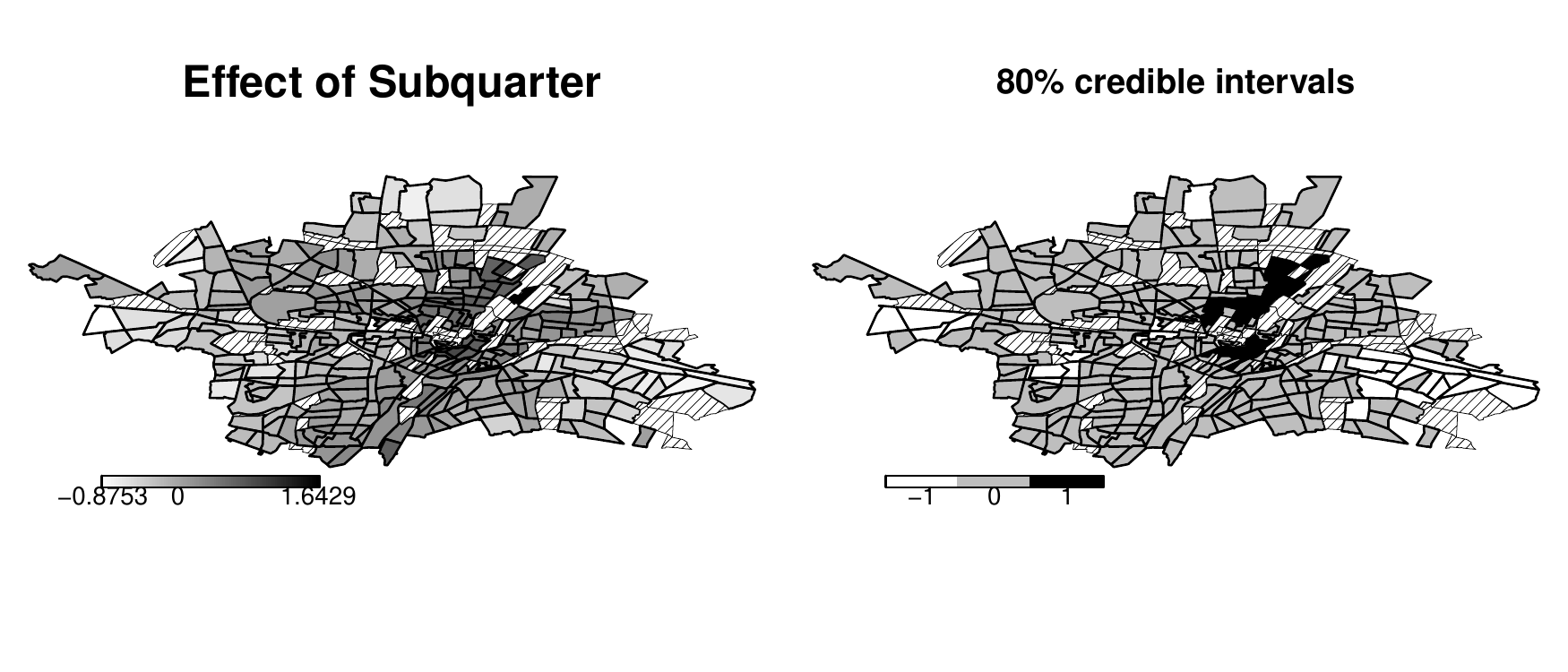}
\caption{\label{mrfEstimates} Map of Munich's subdistricts with
estimated effect on net rent/m$^2$. Left panel shows posterior mean of effects,
right panel shows the sign of 80\% posterior credible intervals: regions with
lower net rent in white, higher net rent in black, regions with credible
intervals overlapping zero in gray. Subdistricts without any
observations are filled with diagonal lines.}
\label{fig:mingaMrfEstimates}
\end{figure}

\begin{figure}[!htbp] \centering
\includegraphics[width=\textwidth]{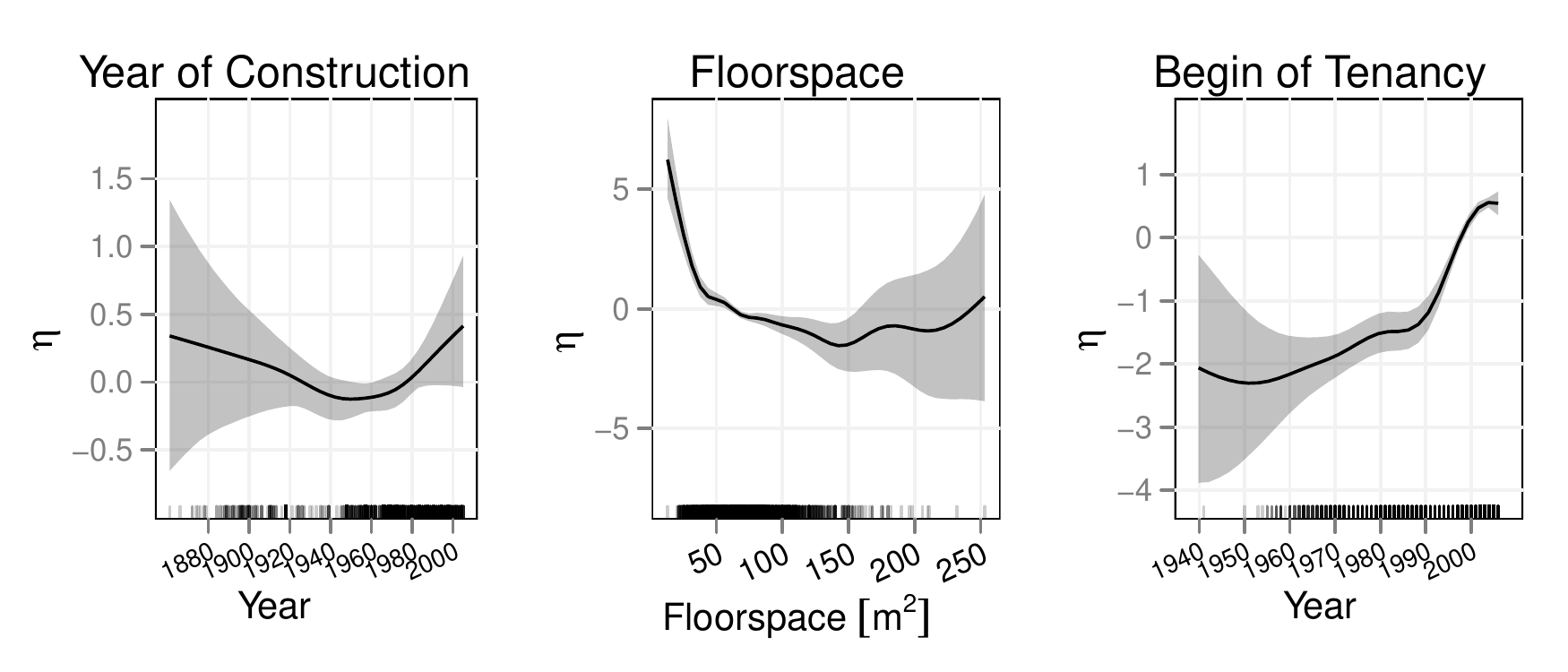}
\caption{Smooth terms with posterior inclusion probability greater
than 10\% for Munich rental guide data. Grey ribbons show 80\% pointwise credible intervals.}
\label{fig:mingaSmooEstimates}
\end{figure}

\begin{table}
\begin{center}
\begin{footnotesize}
\begin{tabular}{lr}
 Term & $P(\gamma=1)$ \\
  \hline
  MRF(subdistrict) & 0.86 \\
  Floorspace, linear & 0.95 \\
  Floorspace, smooth & 0.97 \\
  Year, smooth & 0.10 \\
  Begin of Tenancy, linear & 1.00 \\
  Begin of Tenancy, smooth & 0.97 \\
  Quality of Residential Area, cat. & 0.58 \\
  Company Housing, cat. & 0.82 \\
  Has Planted Area, cat. & 0.26 \\
  Refrigerator, cat. & 0.31 \\
  Kitchen Type, cat. & 0.31 \\
  Intercom, cat. & 0.80 \\
  Flooring, cat. & 0.51 \\
  Has Balcony, cat. & 0.48 \\
  Number of Balconies, cat. & 0.21 \\
  Has Terrace, cat. & 0.34 \\
  Number of Terraces, cat. & 0.15 \\
  Has Roof Terrace, cat. & 0.19 \\
  Number of Roof Terraces, cat. & 0.12 \\
  Area of 2nd Balcony, linear & 0.12 \\
  Has Clothes Drying Area, cat. & 0.15 \\
  Has Playground, cat. & 0.62 \\
  Has Attic, cat. & 0.60 \\
  Garden-Use Permission, cat. & 0.14 \\
  Apartment Type, cat. & 0.15 \\
\end{tabular}
\end{footnotesize}
\caption{Terms with inclusion probability $P(\gamma=1)>0.1$.} \label{T:mingaInclude}
\end{center}
\end{table}
Table \ref{T:mingaInclude} lists the terms with posterior inclusion probability greater than 10\%. The additive predictor is
dominated by the contributions of terms for the presence of balcony, the date of beginning of the tenancy, the quality of the
residential area the property is located in and presence of an attic and presence of a playground. Figures
\ref{fig:mingaMrfEstimates} and \ref{fig:mingaSmooEstimates} show the estimated effect of subdistrict on net rent per square
meter and the continuous predictor effects, respectively, both combined with associated credible intervals. It general, the
estimated effects resemble those found in previously published analyses \citep[cf.~][]{Kneib:Konrath:Fahrmeir:2011} of this
data, with lower rents in predominantly working-class outskirts and higher rents in fashionable districts in and around the
city centre. Including the beginning of tenancy \cite[which has not been included in the analyses
by][]{Kneib:Konrath:Fahrmeir:2011} seems to somewhat attenuate the effect of the year of construction of the flat, and the
total floorspace has a much larger effect on net rent per square meter.


\paragraph{Cross Validation Performance:}

We perform a 10-fold crossvalidation to gain some insight into the stability of the term selection and to compare the
predictive performance of our approach with previous modeling efforts described in \citet{Kneib:Konrath:Fahrmeir:2011}. While
ridge and lasso priors to select single dummy variables of specific factor levels have been used in their model, our results
are based on blockwise selection including or excluding all levels of a categorical covariate simultaneously. Models in
\citet{Kneib:Konrath:Fahrmeir:2011} were estimated with the software package \package{BayesX} \citep{bayesx}. We also
considered ``expert'' models in analogy to \citet{Kneib:Konrath:Fahrmeir:2011} including only a strongly reduced number of
covariates as candidates, estimated both with \package{BayesX} and our function selection approach. For the latter, we also
estimated an expanded ``expert'' model including all potential two-way
interactions (except those with the spatial GMRF term). The resulting
model has 1051 coefficients in 190 model terms including varying coefficient
terms and smooth interaction surfaces.

\begin{figure}[!htbp] \centering
\includegraphics[width=\textwidth]{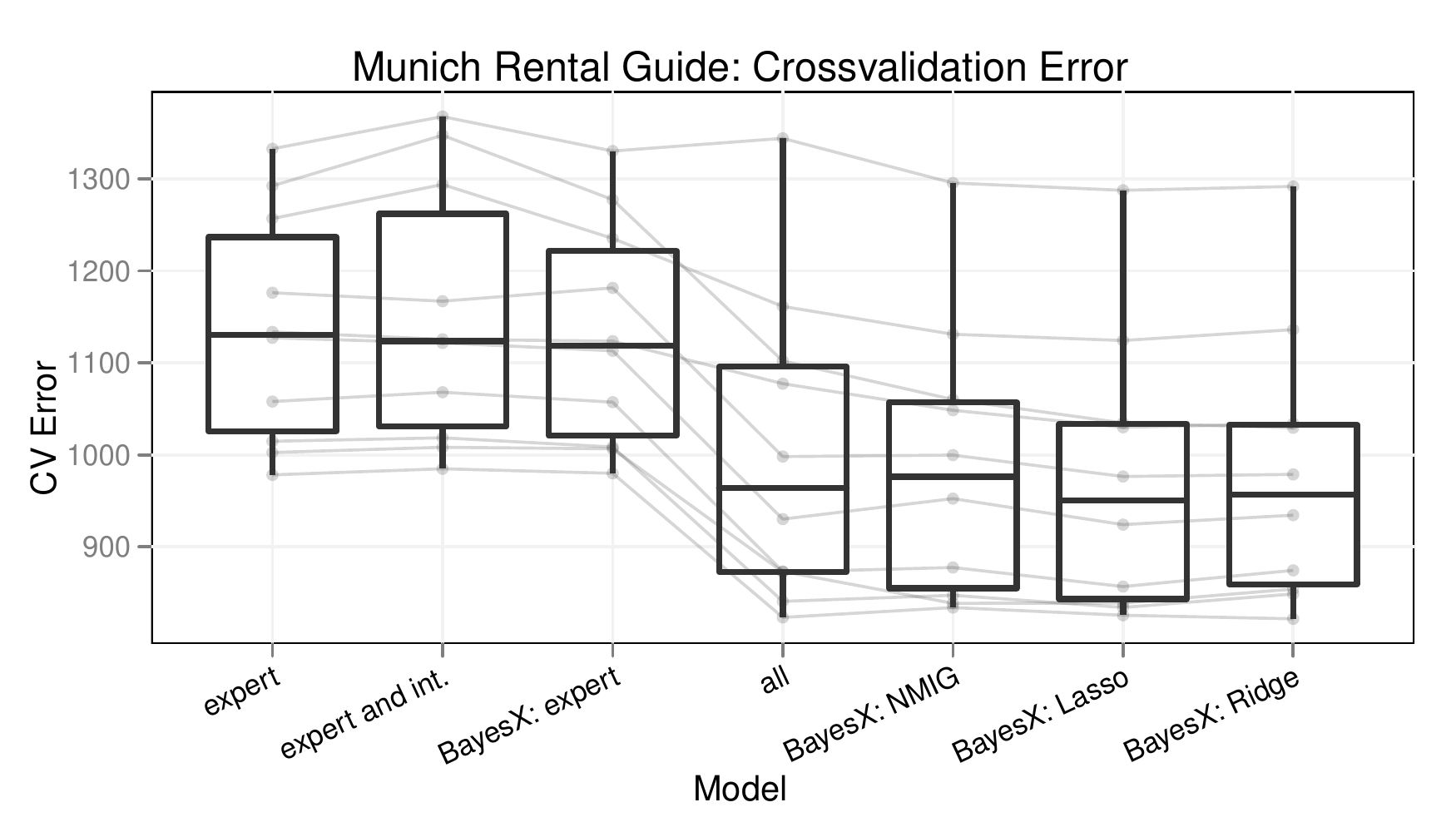}
\caption{Prediction error for 10 cross validation folds for the
Munich rental guide data. Grey lines connect results from identical
folds.} \label{fig:minga_CVError}
\end{figure}

Figure \ref{fig:minga_CVError} displays the cross validation error
on the ten folds for the different model specifications.
We see that prediction accuracy is not diminished by putting a model with only relevant terms under selection, as shown by the
equivalent performances of the peNMIG ``expert''-model and that of the ``expert''-model fit with \package{BayesX} (first and
third boxes from the left). This indicates that the relevant effects are estimated without bias despite the variable selection
prior associated with them, a consequence of the adaptive shrinkage properties of the peNMIG prior. As for the sepsis survival
data analyzed in the subsequent section, there is no noticeable change in prediction performance in most folds if a large
number of interaction terms are added --- compare the similar performances of the expert model and of the expert model with
additional two-way interactions (first and second boxes from the left). The underlying reason is that only a single interaction
effect, the interaction between the level of kitchen furnishing and level of pollution has $P(\gamma=1)>0.1$ in the expanded
``expert'' model, but its effect is still very small. The precision of the prediction achieved by our peNMIG approach on the
full data set with 269 potential model terms (fourth box from the left) is very similar to that of Bayesian lasso, Bayesian
ridge and the conventional NMIG approach (as implemented in \package{BayesX}) for most folds, even though estimates of the
dominating (nonlinear) effects for floorspace, year of construction and begin of tenancy and the effect of subquarter were
associated with a variable selection prior in our model while they where associated with conventional smoothing priors in
\package{BayesX}. This reinforces our conclusion that important effects are estimated without a selection-induced attenuating
bias in our approach.

Variable selection is very stable across the folds, with the same
terms included in all ten folds for the ``expert'' model and the
``expert'' model with interactions. Only
a single one of the interactions (between kitchen furnishings and
level of pollution), has marginal posterior inclusion probabilities
above $0.1$ in six of the ten folds, the rest is excluded
unequivocally in all folds. The model with interactions is more
conservative and includes less of the main effect terms than the
smaller model, because the large number of irrelevant terms moves
most of the posterior mass for $w$, the overall prior inclusion
probability, towards very low values: the posterior means of $w$ in
the smaller model are between $0.71$ and $0.85$, while the posterior
means for $w$ in the model with interactions are between $0.07$ and
$0.09$. In the model with all possible covariates, a core set of 23
covariates is identified in at least nine out of the ten folds,
while nine other covariates have marginal posterior inclusion
probabilities above $0.1$ in at least one fold. Of those nine, two
do so in eight of the folds.

\subsection{Case Study: Hymenoptera Venom Allergy}\label{sec:app:insect}

\paragraph{Data} We reanalyze data on bee and wasp venom allergy from a large observational multicenter study previously analyzed in
\blind{\citet{Rueff:2009}}. The data consists of 962 patients from 14 European study centers with established bee or wasp venom allergy who suffered an allergic reaction after being
stung. The binary outcome of interest is whether patients suffered a severe, life-threatening reaction, defined as anaphylactic shock, loss of consciousness, or cardiopulmonary
arrest. A severe reaction was observed for 206 of the 962 patients (21.4\%). Data were collected on
the concentration of tryptase, a potential biomarker, 
patients' sex and age, whether the culprit insect was a bee or wasp,
on the intake of three types of cardiovascular medication
($\beta$-blockers, ACE inhibitors  and anti-hypertensive drugs),
whether the patient had had at least one minor systemic reaction to a sting
prior to the index sting and the CAP-class (a measure of antibody load) of the
patient with regard to the venom of the culprit insect, with levels $0, 1, 2, 4, 5+$.

\paragraph{Models} An analysis of this data has to take into account possible study
center effects, possible non-linear effects of both age and the
(logarithm of) blood serum tryptase concentrations and the
possibility of differing effect structures for bee and wasp stings.
Our aim is twofold again: We want to (1) estimate a model that allows assessment of the influence
of each covariate on the susceptibility for a severe reaction,
accounting for possibly nonlinear effects and interaction effects
and (2) use this setting to evaluate the stability of the selection
and estimation of increasingly complex models on real data
as well as investigate the consequences of less-than-optimal sampler convergence we observed.

\paragraph{Full Data Analysis}
We fit a peNMIG model with all main effects and all second order interactions
except those with study centre, with smooth functions for both age and tryptase and
a random intercept for the study center. In total, this model has 267 coefficients in 66 model terms:
13 main effects including the global intercept, separate linear and non-linear terms for age and tryptase
and a random intercept for study centre, 21 interactions between the seven factor variables, 28 terms for the
linear and smooth interactions for age and tryptase with each of the seven factors,
and four terms for the interaction effect of age and tryptase
(one linear-linear interaction, two varying coefficient terms, one smooth interaction surface)
Results are based on samples from 20 chains with 40000 iterations each after 1000 burn-in, with every 20$^{th}$ saved.
Running a single chain of this length on a modern desktop computer (i.e., Intel Q9550 2.83GHz) takes about 45 minutes,
so that the entire fit takes about 4 hours on a quad-core CPU.

\begin{figure}[!tbp] \centering
\includegraphics[width=\textwidth]{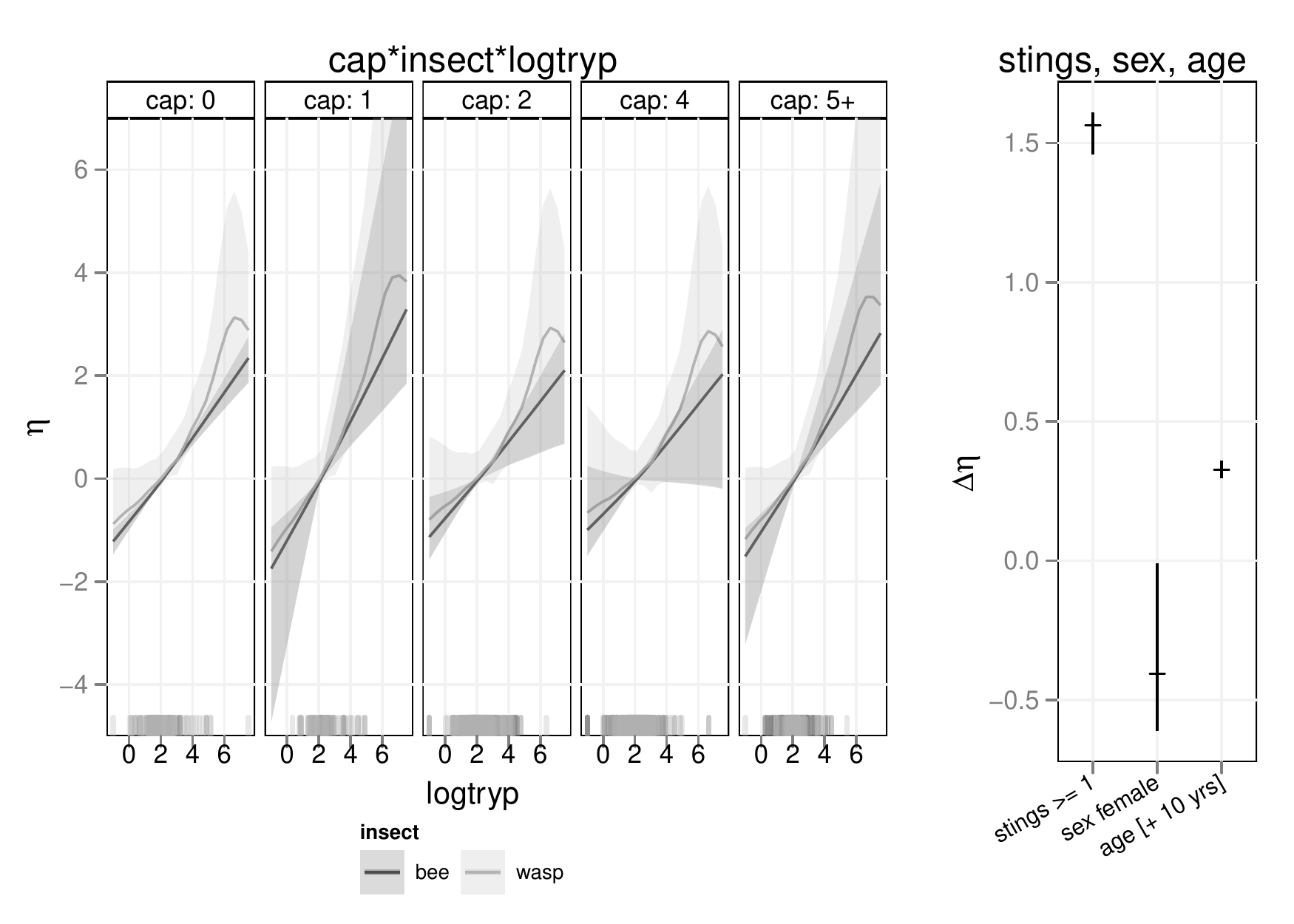}
\caption{Posterior means of effects with (pointwise) 80\% credible intervals. Only effects for terms with marginal inclusion probability $>.1$ are shown.
Since there is some evidence for interlocking interactions of cap class, tryptase and culprit insect
(c.f. Table \ref{tab:insectEffects_All1Long}), the left graph shows the joint effect of these 3 variables.
The graph on the right shows the relative effects of previous stings
(compared to none before the index sting), female gender (compared to male) and an increase in the patient's age by 10 years.}
\label{fig:insectEffects_All1Long}
\end{figure}
\begin{table}[ht]
\begin{small}
\begin{center}
\begin{tabular}{l|c}
 Term & $P(\gamma=1)$ \\
  \hline
  culprit insect & 0.16 \\
  stings & 1.00 \\
  sex & 0.70 \\
  age, linear & 1.00 \\
  tryptase, log-linear & 1.00 \\
  study centre & 0.71 \\
  insect:tryptase, smooth & 0.46 \\
  cap:tryptase, log-linear & 0.14 \\
   \hline
\end{tabular}
\end{center}
\end{small}
\caption{Posterior means of marginal inclusion probabilities $P(\gamma=1)$ (only given for terms with $P(\gamma=1)> .1$).}
\label{tab:insectEffects_All1Long}
\end{table}
Figure \ref{fig:insectEffects_All1Long} shows the estimated effects of the terms with $P(\gamma=1)> .1$ that are listed in
Table \ref{tab:insectEffects_All1Long}. Since the inclusion probabilities indicate interlocking interactions of cap, tryptase
and culprit insect, the panels in the left graph in the figure show the joint effects of these 3 variables. Each panel shows
the effect estimate of tryptase plasma concentration for bee patients (dark grey) and wasp patients (light grey) for the given
CAP class. The rug plot at the bottom indicates the locations of the data. The large uncertainty precludes a detailed
interpretation of this 3-way interaction, but in general, the risk is higher for wasp patients: the main effect of culprit
insect yields an odds ratio of 1.16 (80\%CI: 1-2.43) and the increase in risk in wasp patients seems to be smaller for lower
and larger for higher tryptase concentrations. The graph on the right shows the relative effects of previous stings (compared
to none before the index sting), female gender (compared to male) and an increase in the patient's age by 10 years. Estimated
random effects for the study centres are not shown, their associated posterior mean odds ratios range between 0.44 and 2.13.

\paragraph{Lack of Convergence for $\bm\gamma$}
For this fairly complicated model, we experience some difficulties with the convergence of the MCMC sampler:
We observe poor mixing for some of the entries in $\bm\gamma$, with chains getting stuck in basins of attraction around
posterior modes for long periods of time. This leads to posterior inclusion probabilities for single chains often ending
up either close to zero or close to one for some of the terms.
Running a large number of parallel chains from random starting configurations seems to remedy this problem.
\begin{figure}[!tbp] \centering
\includegraphics[width=\textwidth]{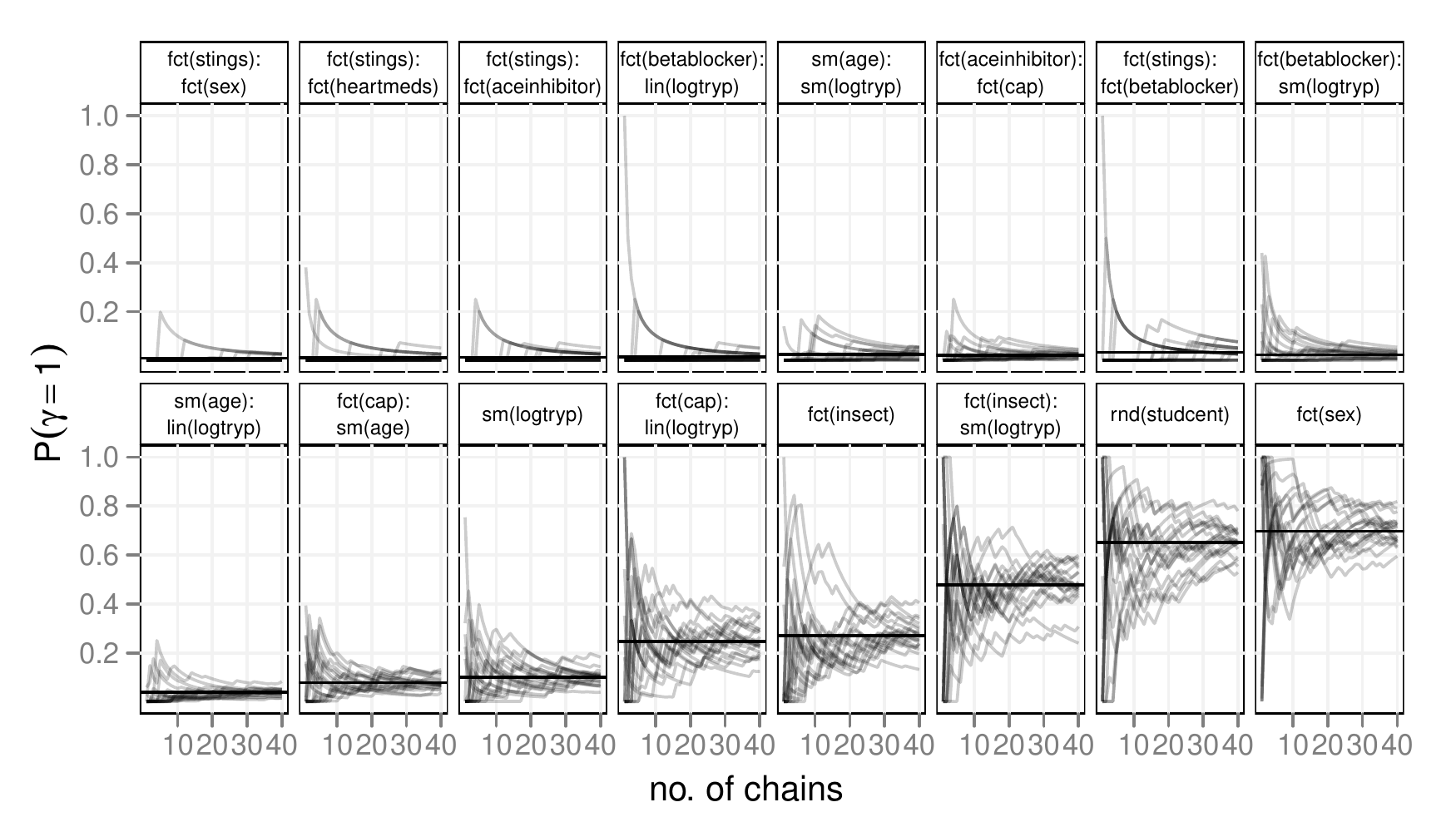}
\caption{Average inclusion probabilites for those terms with convergence issues for 20 fits with 40 chains each.
Grey lines connect posterior means over an increasing number of chains for each fit.
Black horizontal line shows the mean over all 800 chains.}
\label{fig:insectStability}
\end{figure}
To investigate this issue, we perform a large MCMC experiment with
800 chains, each with 10000 iterations after 100 burnin, for the
model described above. Figure \ref{fig:insectStability} shows the
average inclusion probabilites for the 16 terms with the highest
between-chain variability of $P(\gamma=1)$ for 20 fits with 40
chains each. Grey lines connect posterior means based on an
increasing number of chains for each fit. The black horizontal line
shows the mean over all 800 chains, which we presume to be a good
estimate of the ``true'' marginal posterior inclusion probability.
Convergence of the posterior means is slow for these terms, but
discrimination between important, intermediate and negligible
effects seems to be reliable based on as few as 10 to 20 chains.
While we would not be comfortable in claiming that 10 or 20 parallel
chains are enough to completely explore this very high-dimensional
model space and yield a reliable estimate of posterior model
probabilities, i.e., the joint distribution of $\bm\gamma$, the
marginal inclusion probabilities $P(\gamma_j=1)$, $j=1,\dots,p$ of
the various terms seem to be estimated well enough to distinguish
between important, intermediate and negligible effects, which is
usually all that is required in practice. This conclusion is also
borne out by the UCI benchmark study in the main article.

\paragraph{Predictive Performance Comparison}
We subsample the data 20 times to construct independent training
data sets with 866 subjects each and test data sets with the
remaining 96 patients to evaluate the precision of the resulting
predictions and compare predictive performance to that of equivalent
component-wise boosting models fitted with \package{mboost} and an
unregularized GAMM-fit with all main effects estimated with
\package{gamm4}. Results for our approach are based on 8 parallel
chains each running for 10000 iterations after 500 iterations of
burn-in, with every 10$^{th}$ iteration saved. Component-wise
boosting results are based on a stopping parameter determined by a
25-fold bootstrap of the training data, with a maximal iteration
number of 500. We compare three model specification of increasing
complexity: a simple model with main effects only, a model with main
effects and all interactions between culprit insect and the other
covariates, and the complex model with all main effects and second
order interactions presented before. We were unable to fit the
latter model with \package{mboost}, and the model including the
insect interactions could not be fitted by \package{mboost} for 4 of
the training data sets. We report results for the 16 sets remaining.
\begin{figure}[!tbp] \centering
\includegraphics[width=\textwidth]{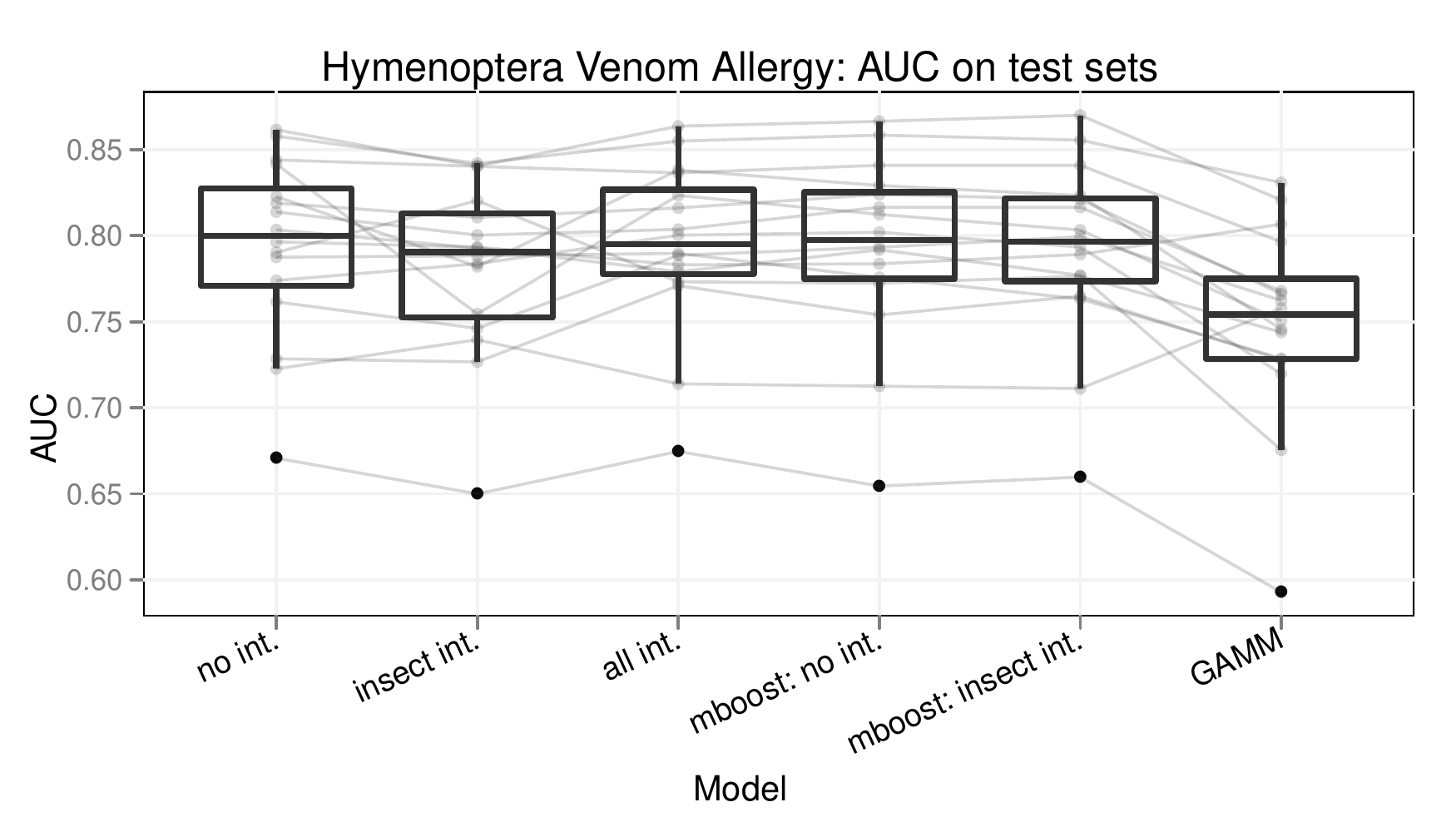}
\caption{Area under the ROC curve for 20 test sets from the hymenoptera venom allergy data set, higher is better.
Grey lines connect results from identical folds. }
\label{fig:insect_CVAUC}
\end{figure}
Figure \ref{fig:insect_CVAUC} shows the area under the ROC cuve (AUC) achieved by the different model specifications.
For this data set, the models with higher maximal complexity show slight decreases in predictive accuracy,
but still perform better than an unregularized generalized additive mixed model (GAMM) on the far right.

Despite the fairly low number of parallel chains and comparatively short chain lengths, the stability of the marginal term
inclusion probabilities across subsamples is fairly good, indicating that the term selection is robust to small changes in the
data and that even as few as 8 chains may be enough to reach fairly reliable rough estimates of term importance in this
difficult setting. All model specifications identified the same subset of important main effects (i.e., number of previous
stings before the index sting, sex, linear effects of age and the log of tryptase and the random effect for study centre).
\begin{figure}[!tbp] \centering
\includegraphics[height=.9\textheight]{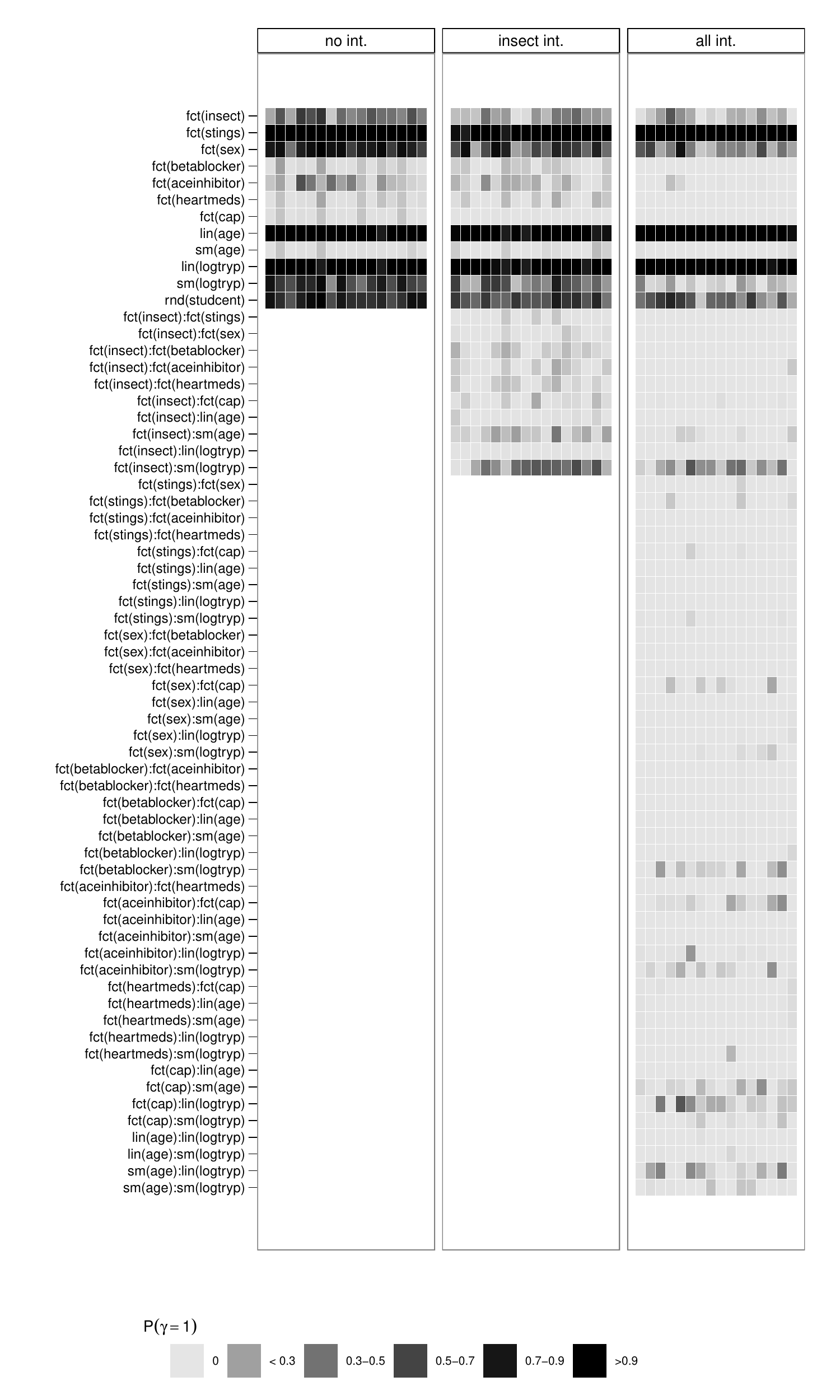}
\caption{Posterior means of inclusion probabilities $P(\gamma=1)$ across 16 subsampled training data sets for the 3 model specifications.}
\label{fig:insect_CVIncProb}
\end{figure}
Figure \ref{fig:insect_CVIncProb} shows the posterior means of inclusion probabilities $P(\gamma=1)$ across 16 subsampled
training data sets for each of the 3 model specifications.

\clearpage
\begin{small}
\bibliographystyle{Chicago}
\bibliography{spikeSlabGAM}
\end{small}

\end{document}